\documentclass[12pt]{article}

\usepackage{newtxtext,newtxmath}

\usepackage{amsthm}

\newtheorem{theorem}{Theorem}[section]

\newtheorem{conjecture}{Conjecture}[section]
\usepackage{graphicx}
\usepackage{tikz}
\usetikzlibrary{positioning, arrows.meta}
\usepackage{tcolorbox}
\usepackage{subcaption}
\usepackage{xcolor}

\definecolor{node_red}{RGB}{255, 153, 153}
\definecolor{node_blue}{RGB}{153, 204, 255}

\usepackage[letterpaper,margin=1in]{geometry}

\renewenvironment{abstract}
	{\quotation}
	{\endquotation}

\date{}

\makeatletter
\renewcommand{\fnum@figure}{\textbf{Figure \thefigure}}
\renewcommand{\fnum@table}{\textbf{Table \thetable}}
\makeatother

\usepackage{scicite}

\usepackage{url}

\def\scititle{Symbiosis as a systemic catalyst and the impossibility of coalitions in optimal networks}
\title{\bfseries \boldmath \scititle}

\author{
	Giulia~Palma$^{1}$,
	Antonio~Rizzo$^{1}$,
	Chiara~Mocenni$^{2\ast}$\and
	\small$^{1}$Department of Social, Political and Cognitive Sciences, University of Siena, Siena \& 53100, Italia.\and
	\small$^{2}$Department of Information Engineering and Mathematics, University of Siena, Siena \& 53100, Italia.\and
	\small$^\ast$Corresponding author. Email: chiara.mocenni@unisi.it\and
}

\begin{document} 

\maketitle

\begin{abstract} \bfseries \boldmath 
The stability of complex systems hinges on the tension between individual incentives and collective welfare. Modeling these dynamics through strategic network interactions based on anti-coordination, we formally prove that any globally optimal configuration constitutes a Strong Nash Equilibrium, creating topological barriers against collective deviations. However, in sub-optimal states, strictly individualistic agents remain trapped in stagnant equilibria. We show that coalition formation acts as a vital catalyst for global efficiency. Paralleling Tomasello?s evolutionary theory of shared intentionality, the emergence of symbiotic joint agency overcomes selfish stagnation and drives the system toward optimal niche partitioning. We validate our framework through extensive computational simulations and apply it to an empirical pollination network, demonstrating how symbiosis may steer real-world ecosystems toward maximum resilience. We uncover metastable dynamics where coalitions continuously reconfigure, revealing that biological evolution relies on a perpetual, adaptive balance between competition and cooperation.
\end{abstract}

\section*{Introduction}

The stability and resilience of complex systems emerge from a delicate and continuous trade-off between individual (microscopic) incentives and collective (macroscopic) welfare \cite{Newman2003, JUSUP20221}. In a broad class of dynamical systems, interactions are governed by anti-coordination dynamics: scenarios where a single agent's utility increases upon adopting strategies or traits that diverge from those of its neighbors, a mechanism elegantly described by potential games \cite{Monderer1996}. Recent developments in social physics and equilibrium convergence highlight how these conflicts of interest can actually improve collective computation and drive self-organization \cite{Pangallo2019, Brush2018}. Understanding how purely local and selfish decisions can guide an entire system toward globally optimal configurations represents a fundamental challenge in the theory of complex adaptive systems \cite{Levin2003}.

Anti-coordination dynamics and the drive for differentiation are the engines of countless natural phenomena \cite{Tilman1994, Bascompte2009}. In ecology, competition for spatial and nutritional resources forces species toward ecological niche partitioning; this continuous evolutionary pressure to differentiate from competitors promotes coexistence, protects against coextinctions, and ensures the long-term stability of biodiversity, especially in complex bipartite networks such as plant-pollinator mutualisms \cite{McCann2000, Montoya2006, Adler2018, Barabas2016, Chen2024, Morton2022, THEBAULT2020, ARROYO1985}. Such interactions structure the intricate web of life, mitigating delay effects in population dynamics \cite{Pigani2022} and shaping evolutionary responses, such as the emergence of mutualistic hotspots or bacteria-phage survival dynamics \cite{Lawrence2012, Ametrano2021, Wright2016, Cosmo2024, McPeek2025}. Similar phenomena were crucial in major evolutionary transitions, enabling the emergence of cooperation and altruism \cite{Szathmary1995, Nowak2006, West2007, Lean2022}. Similarly, in the socio-economic sciences, network formation and spatial rock-paper-scissors games demonstrate how anti-coordination decisively shapes the structure and efficiency of the social fabric \cite{Epstein1999, AntiCoordination, RPSNetworks, RPS2014, Tu2018, Impossible2023, Akbarpour2018}. Moreover, several methods to contrast polarization have been studied by developing suitable adaptive strategies \cite{polarization} organic
visualization  to connect people of opposing views on social networks using \cite{GraellsGarrido2013DataPC}.

From a formal standpoint, the mathematical quintessence of anti-coordination interactions is captured by the max $k$-cut problem, also known as the strategic graph coloring game \cite{max_cut_encyclopedia, all-or-nothing, gourves-monnot-3, generalized}. In this formulation, nodes strategically choose a color from available alternatives to maximize connections with heterogeneous neighbors, a mechanism also explored in defective coloring or domination coloring variants \cite{Defcolor1, Defcolor2, dom_col_graph}. Although finding the exact global optimum for this partitioning is a classic NP-complete problem -- often tackled with memetic approaches or complex global optimization algorithms \cite{memes, vilmar} -- the game-theoretic perspective, which has also been extended to hypergraphs or sub-optimal states \cite{smorodinskis, mt}, allows us to investigate the intrinsic tension between individual greed and network efficiency.

The abstraction of the max $k$-cut problem provides crucial architectural tools to solve real-world challenges. In telecommunications and systems engineering, this approach is employed for optimizing wireless networks and frequency allocation using quantized conflict graphs \cite{wireless}. In computational social sciences, architectures based on this principle are used to mitigate social polarization, redesigning connections to foster dialogue between individuals with opposing views and fragmenting echo chambers \cite{opposite_opinions}.
However, a fundamental question remains: does a system strive for a final, frozen state of perfection, or is its vitality found in the journey itself? We argue that while the global optimum represents a theoretical 'climax' of absolute stability, the actual life of a complex network is a sequence of strategic shifts -- a perpetual dance of coalitions that keeps the system in a state of productive metastability.

In this study, we resolve the paradox between individual greed and collective stability. We demonstrate that when a system is developing and sub-optimal, the structured formation of coalitions acts as a catalyst for global efficiency, a phenomenon consistent with symbiotic dynamics in eco-industrial parks or artificial intelligence \cite{Zheng2021, Bistaffa2021}. However, we provide a rigorous formal proof that every globally optimal configuration in the max $k$-cut game constitutes a Strong Nash Equilibrium (SE) \cite{gascom, MDPI2022, MDPI2024}. Expanding upon the classical literature on strong equilibria in cooperative and network games \cite{aumann-se-1, aumann-se-2, gourves-monnot-1, gourves-monnot-2, Apt2008, Feldman2009, Chalkiadakis2010}, we prove that in a state of maximum social welfare, no collective deviation is possible. In a fully mature selfish society, the network architecture itself acts as a topological barrier, rendering opportunistic symbiosis mathematically impossible and thus guaranteeing the incorruptibility and resilience against catastrophic shifts in the ecosystem \cite{Allesina2011, Scheffer2001}.
This leads to a structural paradox: the very architecture that guarantees immunity against opportunistic deviations in optimal states also imposes a total cessation of strategic dynamics. By proving the existence of a 'topological barrier' in optimal networks, we highlight that the vitality of an ecosystem resides not in its final climax, but in the sequence of sub-optimal, metastable configurations where symbiosis can still act as a transformative catalyst.

\section{Problem Formulation: The Mechanics of Collective Variation}

To rigorously examine the boundaries of collective stability, we first establish the mathematical framework of the max $k$-cut game and its equilibrium properties. 
This game is played on an undirected, unweighted graph $G = (\mathcal{V}, \mathcal{E})$, where the set of $n$ vertices $\mathcal{V} = \{1, \dots, N\}$ represents a population of autonomous agents -- be they individuals in a social network or species in an ecosystem. In such systems, 
the topological structure of the network is a primary driver of the emergent behavior and functional dynamics of the population \cite{Newman2003}. The set of edges $\mathcal{E}$ defines the interconnecting substrate, representing, for example, competitive interactions among members of the population.

The strategic essence of the game lies in the choice of a color (or trait) from a finite set $\mathcal{K} = \{1, \dots, M\}$. A strategy profile, or coloring $\sigma = (\sigma_1, \sigma_2, \dots, \sigma_n)$, assigns a specific trait to each agent. 
Each coloring $\sigma$ induces a partition of the vertices into $k$ sets; the \textit{cut} of the graph is thus the set of edges that connect nodes belonging to different partitions (i.e., nodes with different colors). 
In the max $k$-cut game, agents are inherently motivated by anti-coordination: the individual utility, or payoff $\mu_v(\sigma)$, of a player $v$ is the count of its neighbors that have chosen a different color. Mathematically, this is expressed as:
\begin{equation}
\mu_v(\sigma) = \sum_{w \in \mathcal{V}, \sigma_w \neq \sigma_v} a_{v,w}, \label{node_payoff}
\end{equation}
where $a_{v,w}$ are the entries of the adjacency matrix $A$. Considering, by assumption, unweighted graphs, this matrix is symmetric, as it is the interaction between pairs of individuals or species: competition from only one side is unilateral exploitation. 
Anyway, the framework or the present study can be extended to weighted graphs where the entries $a_{v,w}$ represent the intensity of competitive pressure or the 
magnitude of resource overlap between agents, providing a more granular representation of competitive and coordinated tensions in networked populations \cite{Apt2008}.

The significance of the payoff function described by equation \eqref{node_payoff} from an ecological standpoint relies on niche partitioning, where species maximize their fitness by differentiating phenotypic traits to avoid direct overlap and competition for the same resources to ensure stable coexistence in high-diversity environments, to maintain sufficient functional differences and minimize competitive exclusion \cite{Tilman1994}. Furthermore, ecological theory suggests that stable coexistence is maintained because intraspecific competition is significantly stronger than 
interspecific competition \cite{Adler2018}, effectively penalizing agents that occupy identical niches and reinforcing the global stability of the system.

The aggregate performance of the system is measured by the social welfare $\mu(\mathcal{V},\sigma)$, defined as the sum of all individual payoffs:
\begin{equation}
\mu(\mathcal{V},\sigma) = \sum_{v \in \mathcal{V}} \mu_v(\sigma).\label{social}
\end{equation}
 In combinatorial terms, $\mu(\mathcal{V},\sigma)$ is exactly twice the value of the cut induced by the partition of colors. A configuration that achieves the maximum possible social welfare is termed an Optimal Coloring (or Global Optimum). Such a state represents a biological climax where the total tension or frustration in the network is minimized, and resource distribution is globally efficient.
The transition from individual incentives to collective stability is captured by the hierarchy of equilibrium concepts. An agent-based system is in a Nash Equilibrium (NE) if no single player can improve their payoff by unilaterally changing their color. While NE ensures stability against individual greed, it does not account for coordinated actions. To address this, we employ the concept of Strong Equilibrium (SE). Let $C\subseteq \mathcal{V}$ be the set of nodes that can increase their payoff, thus being the candidates to form a coalition. In the following, we name the coalitions by $F$, where $F \subseteq{C} \subseteq{\mathcal{V}}$. Moreover, we define the inertial set $D=C\setminus F$ composed of $C$ that does not enter any coalition. A coloring $\sigma$ is an SE if there exists no coalition $F$ that can cooperatively shift to a new coloring $\gamma$ such that any of its members strictly improves their individual utility:
\begin{equation}
\Delta\mu_v(\gamma, \sigma) = \mu_v(\gamma) - \mu_v(\sigma) > 0, \quad \forall v \in C.
\end{equation}
Analogously, the social welfare variation, $\Delta\mu(\gamma, \sigma) = \mu(\gamma) - \mu(\sigma)$, is strictly less than zero. 

In our framework, such a strong deviation is synonymous with the emergence of symbiosis: a group of agents coordinating their traits to bypass local sub-optimality. This process mirrors the major transitions in evolution, where previously independent entities merge into higher-level functional units to overcome constraints on efficiency and information transmission \cite{Szathmary1995}. 

The central question of this work -- and a long-standing conjecture in the field -- is whether a globally optimal state is inherently an SE, thereby acting as a topological barrier that renders the spontaneous emergence of new symbiotic coalitions impossible once the system has reached its maximum efficiency.

To visualize the role of symbiosis as an evolutionary catalyst, we present a schematic comparison in Box 1; this example demonstrates that while individual agents may remain stagnant in a suboptimal equilibrium, a coordinated coalition can unlock a transition toward the biological climax.
The Diamond Graph example illustrates the fundamental limitation of individual strategic choices. In the coloring $\sigma$, node $v_2$ is trapped in a suboptimal state: a unilateral change of color would result in no net payoff improvement ($\Delta \mu_{v_2} = 0$), defining $\sigma$ as a Nash Equilibrium. However, this state is not socially efficient. By forming a coordinated coalition $F = C = \{v_2, v_3\}$, the agents can overcome this local impasse. Through a synchronized strategy shift (symbiosis), both members strictly improve their condition, allowing the system to reach the Global Optimum $\gamma$. This transition underscores that while individuals may remain stagnant in suboptimal equilibria, the emergence of collective coordination acts as a catalyst for systemic efficiency and biological climax.
\begin{figure}[ht!]
\begin{tcolorbox}[colback=white, colframe=black!70, title=Box 1: From Nash Trap to Global Optimum, fonttitle=\bfseries]
    \centering
    \begin{minipage}{0.45\textwidth}
        \centering
        \textbf{Nash Equilibrium ($\sigma$)}\\[1ex]
        \begin{tikzpicture}[node distance=1.5cm, every node/.style={circle, draw, minimum size=0.7cm, font=\small}]
        
        \node[fill=node_red] (v1) {$v_1$};
        
        \node[fill=node_red, right=of v1] (v2) {$v_2$};
        
        \node[fill=node_blue, below=of v2] (v3) {$v_3$};
        
        \node[fill=node_blue, below=of v1] (v4) {$v_4$};
            
        \draw (v1) -- (v2);
        \draw (v2) -- (v3);
        \draw (v3) -- (v4);
        \draw (v4) -- (v1);
        \draw (v1) -- (v3);
        \end{tikzpicture}
    \end{minipage}
    \hfill
    \begin{minipage}{0.45\textwidth}
        \centering
        \textbf{Global Optimum ($\gamma$)}\\[1ex]
        \begin{tikzpicture}[node distance=1.5cm, every node/.style={circle, draw, minimum size=0.7cm, font=\small}]
        \node[fill=node_red] (v1) {$v_1$};
        \node[fill=node_blue, right=of v1] (v2) {$v_2$};
        \node[fill=node_red, below=of v2] (v3) {$v_3$};
        \node[fill=node_blue, below=of v1] (v4) {$v_4$};
    
        \draw (v1) -- (v2);
        \draw (v2) -- (v3);
        \draw (v3) -- (v4);
        \draw (v4) -- (v1);
        \draw (v1) -- (v3);
        \end{tikzpicture}
    \end{minipage}
    \vspace{3mm}
    \begin{tcolorbox}[colback=gray!5, colframe=white, size=small]
        \small
        \begin{itemize}
            \item \textbf{Local Conflict:} In $\sigma$, the edges $(v_1, v_2)$ and $(v_3, v_4)$ are not cut as they have the same color.
            \item \textbf{Individual Stability:} In configuration $\sigma$, no single agent can strictly improve their payoff through a unilateral color change, thus confirming they are in a Nash equilibrium.
            \item \textbf{Coalition:} $F = \{v_2, v_3\}$ is composed by agents who collectively switch strategy.
            \item \textbf{Social Welfare Improvement:} $\mu(\mathcal{V}, \sigma) = 6 \longrightarrow \mu(\mathcal{V}, \gamma) = 8$.
        \end{itemize}
    \end{tcolorbox}
\end{tcolorbox}

%
\label{fig:nash_optimum_box}
\end{figure}

\section{Preliminary Results}\label{preliminary}

For the purpose of our analysis, we assume the population has already reached a sub-optimal configuration, such as a Nash equilibrium. In these states, individual agents are unable to improve their fitness unilaterally; however, the formation of coordinated coalitions allows them to bypass this stagnation, jointly increasing their own payoffs and driving the entire system toward a more efficient state.

To rigorously analyze this transition, we establish a fundamental bridge between individual utilities and group-level topological features, accounting for the reciprocal nature of interactions within a coalition. This decomposition is crucial for isolating the internal synergy of a group from its external interactions with the rest of the network. As suggested by the definition of payoff in equation \ref{node_payoff}, the aggregate payoff of a group is determined by the network topology: it equals the sum of all cut edges connecting the group to the external network, plus twice the number of internal cut edges.

The resilience of global optima is rooted in the structural composition of the utility earned by a coalition. According to the definition of payoff of a single node, reported in equation \eqref{node_payoff}, the aggregate payoff of a group is determined by the network topology: it equals the sum of all cut edges connecting the group to the external network, plus twice the number of internal cut edges. This double-counting reflects the reciprocal nature of the max $k$-cut game, where an internal edge contributes to the fitness of both participants, thereby characterizing the synergistic potential of collective behavior.

Building upon this, we  derive an identity governing the variation of global social welfare during coordinated deviations. The net change in the total system cut is intrinsically linked to the sum of individual payoff improvements, corrected by the change in internal cut edges. In  mathematical terms, the global variation equals twice the sum of individual improvements minus twice the net growth of internal cut edges. This identity bridges micro-level motives and macro-level outcomes, revealing that a group's selfish drive is inextricably tied to the global potential of the interaction graph.

A pivotal implication is the monotonic nature of the game. Since the max $k$-cut is an exact potential game \cite{Monderer1996}, any strong deviation -- where every member strictly improves their utility -- must result in a strictly positive increase in global social welfare. In the context of social physics and biodiversity, this ensures that symbiotic guilds act as directional catalysts, dragging the population toward higher states of collective efficiency until a social optimum is achieved.

However, this capacity for improvement is subject to a strict topological ceiling. A coalition's ability to increase its internal cut is limited by the initial conflicts(edges between same-colored nodes) within the group. We find that the surplus of internal harmony generated during a move can never exceed the total conflict initially broken. Crucially, in a globally optimal configuration, where conflicts are already minimized, the potential for further internal improvement is mathematically constrained to be less than the size of the coalition itself. This provides the formal justification for why globally efficient systems are resilient against fragmentation: the benefits reachable through coordination are insufficient to overcome the stability of the pre-existing global equilibrium.

All details and formal proofs of the preliminary findings reported in Section \ref{preliminary} and the main results discussed in Section \ref{symbiosis} are extensively presented in the Supplementary Material document.

\section{Main Results: Resilience and Symbiotic Catalysis}\label{symbiosis}

The central contribution of this work is the formalization of the link between global optimization and collective stability. We demonstrate that when a system reaches its maximum social welfare, it becomes impervious to coordinated attempts at deviation, regardless of the coalition's size or strategic complexity.

\begin{theorem}[Optimality and Collective Resilience]
In any unweighted and undirected max $k$-cut game, every globally optimal coloring $\sigma^*$ is a Strong Equilibrium (SE).\label{thm:optimality}
\end{theorem}

The formal proof of Theorem \ref{thm:optimality} is provided in the Supplementary Material and proceeds by contradiction. We show that if a coalition were to successfully execute a strong deviation thus improving every member's utility, the strict topological bounds governing the internal edges of the group would force a strictly positive increase in the global cut. This creates a logical paradox, as $\sigma^*$ is already defined as the state of absolute maximum welfare. 

Theorem \ref{thm:optimality} establishes the global optimum as a structural trap: the network architecture itself acts as a topological barrier, preventing any group of selfish agents from collectively escaping a state of maximum social efficiency. In this sense, it identifies the global optimum as a 'frozen' state. Once the system reaches maximum welfare, the network's topology locks every agent into a configuration where no collective movement is profitable. 

This rigidity identifies the global optimum as a 'frozen' state. However, the path toward such a climax is often obstructed by metastable traps---sub-optimal Nash Equilibria that are robust against individual moves but vulnerable to collective ones. To quantify this resistance, we define the metastability radius $m$ as the minimum critical mass of agents required to form a functional symbiotic coalition capable of breaking the local equilibrium. As formally derived in the Supplementary Material, a configuration is said to be $m$-rank resilient if any group of size $|F| < m$ is structurally unable to improve its collective standing. This parameter $m$ represents the energy barrier of the network topology, ensuring that systemic evolution proceeds through discrete, coordinated jumps rather than continuous individualistic shifts.

Anyway, the formally proven 'rigidity' of the optimum serves as a baseline to understand why real-world systems often linger in metastable states: the cost of breaking the final 'welding points' of conflict is high enough to preserve a dynamic, rather than static, equilibrium. Beyond the stability of the climax state, we investigate the dynamics of evolving, sub-optimal populations. A crucial observation in this context is the behavior of the inertial set $D$, composed of candidate nodes that possess an incentive to move but do not participate in a specific coalition's deviation. We establish two results regarding the nodes of the boundary set $D$. First of all, we find a lower bound of the payoff losses to which these nodes undergo as a consequence of the coalition, which is given by the dimension of the coalition itself ($|F|$) minus 2.  

Additionally, we prove that under the hypothesis that the set of colors taken by the coalition and the set of colors of the boundary set $D$ are disjoint, the coalition can produce a beneficial effect on the whole system as it is shown that the difference in the global payoff is strictly positive also in the boundary set $D$. As a consequence, considering that any member of the strong coalition also improves their payoff, and that the nodes of the external set $E$ are resilient to any change induced by the coalition, the system welfare is strictly improves. 

This phenomenon suggests that symbiotic coordination does not merely serve its members, but it may also generate positive externalities by clearing saturated ecological niches and reducing competitive pressure on neighboring species. Within the framework of evolutionary social semantics, such coordinated shifts represent a form of mutual benefit where strategic adjustments by a group enhance the fitness of non-participating agents \cite{West2007}. 

The previous findings on the effect of the strong coalitions on the social welfare suggest extensions to the general case, thus providing a theoretical framework to overcome the individualistic Nash traps. Under the effect of coalitions, sub-optimal would be allowed to move from one static state to a richer metastable one driving the whole system towards effective welfare  improvements. Since a formal proof of this issue is not straightforward using the methodological framework of this paper, we formulate the following conjecture.

\begin{conjecture}[Monotone Social Welfare]
\label{thm:monotone}
Any strong deviation by a coalition $F$ from a sub-optimal (Nash) coloring strictly increases the  social welfare of the system ($\Delta \mu(\gamma, \sigma) > 0$).
\end{conjecture}

Although we do not provide a formal proof of the conjecture in the general case, and without specific assumptions, in the next section, we report the results of an extensive experimental analysis we conducted with the aim of finding any counterexamples. Specifically, we investigate, by means of  combinatorial search and greedy algorithms, whether any strong coalition formed by a Nash color configuration may strictly improve members' payoff and, at the same time, induce a payoff decreasing in the members of the boundary set $D$, thus leading  the global welfare to reduce.  





\section{Experimental Results and Applications}
To evaluate the practical implications of our formal proofs and assess the robustness of the Strong Equilibrium in dynamic environments, we complement our theoretical analysis with a two-fold validation and two applications. 

The section is organized as follows: first, in subsection \ref{greedy} we present a computational validation of conjecture \ref{thm:monotone}, for which we have a monotone payoff increase due to the effects of the strong coalitions grounded on Nash equilibrium.  Moreover, in subsection \ref{abmse}, we utilize agent-based modeling (ABM) to investigate the dynamical pathways through which symbiotic coalitions drive synthetic networks toward higher states of social welfare, thus confirming conjecture \ref{thm:monotone}.  Subsequently, in subsection \ref{bio}, we apply this framework to a real-world ecological substrate -- a high-altitude Andean pollination network --  tracing its evolutionary transition from ancestral taxonomic frustration to a resilient, optimized functional climax.  Finally, in subsection \ref{tomasello}, we bridge the gap between network topology and evolutionary anthropology, framing the transition from individualistic stagnation to collective efficiency as a structural precursor to the emergence of shared intentionality in the human lineage. Together, these results bridge the gap between abstract combinatorial properties, the self-organizing mechanics of complex systems, with a particular focus on the biological ones, and the cognitive architectures of cooperation.

\subsection{Computational validation of the monotone increase in social welfare induced by coalitions}\label{greedy}

To complement the analytical proof and assess the robustness of the Systemic Altruism result, we conducted an independent computational validation via exhaustive enumeration and statistical sampling. 

The objective was to determine whether in any sub-optimal Nash configuration $\sigma$, a coalition $F$ performing a strong deviation to $\gamma$, and a non-empty inertial set $D$, could produce a strictly negative aggregate payoff variation ($\sum_{w} \Delta \mu_w < 0$), thus constituting a counterexample.

The search was organized in two phases. In the first phase, we performed exhaustive enumeration over all connected, unweighted, undirected graphs with $N \leq 5$ nodes and $k \in \{3, 4\}$ colors. For each graph, all possible colorings were tested for the Nash Equilibrium condition, and every valid NE configuration was subjected to a complete search over all coalitions of size $2 \leq F \leq N - 1$, and all possible strong deviations. This phase examined 690 NE configurations for $N = 4$, $k = 3$, 3,156 for $N = 4$, $k = 4$, 23,460 for $N = 5$, $k = 3$, and 138,900 for $N = 5$, $k = 4$, totaling over 166,000 NE configurations under full enumeration.

In the second phase, we extended the search to larger graphs via random sampling. For each combination $(n, k)$ with $N \in \{6, 7, 8\}$ and $k \in \{3, 4\}$, we generated 10,000 random connected graphs and sampled 20 random colorings per graph, retaining those satisfying the NE condition. This phase examined an additional 83,000 NE configurations across
graph sizes up to $N = 8$.

Across all phases, no counterexample was found. Every strong deviation from a Nash Equilibrium produced a non-negative aggregate payoff variation for the inertial set $D$. The complete results are summarized in Table \ref{tab:monotone}.

\begin{table}[h]
\centering
\caption{Counterexamples found in the developed experiments.}
\label{tab:monotone}
\begin{tabular}{|ccccc|}
\hline
$N$ & $k$ & Method & NE Configurations Tested & Counterexamples \\ \hline
4 & 3 & Exhaustive & 690 & 0 \\
4 & 4 & Exhaustive & 3,156 & 0\\
5 & 3 & Exhaustive & 23,460 & 0\\
5 & 4 & Exhaustive & 138,900 & 0\\
6 & 3 & Sampling & 15,309 & 0\\
6 & 4 & Sampling & 16,511 & 0\\
7 & 3 & Sampling & 9,055 & 0\\
7 & 4 & Sampling & 8,265 & 0\\
8 & 3 & Sampling & 5,271 & 0\\
8 & 4 & Sampling &4,039 & 0\\
 \hline
\end{tabular}
\end{table}

This computational evidence strongly supports the validity of payoff monotone increase beyond the range covered by the analytical proof. We note that in the above computational experiments, $\Delta(F,D) \geq 0$ holds without additional assumptions on the deviation structure, which constitutes an open problem of independent combinatorial interest. The exhaustive verification for $N \leq 5$ and the absence of counterexamples under extensive sampling for $N \leq 8$ provide robust empirical grounding for the result as stated.

\subsection{Simulation Experiments and Agent-Based Modeling}
\label{abmse}

To validate our theoretical framework, we implemented an agent-based model (ABM) simulating a system of $N=200$ selfish agents. Following the generative
approach to social and biological science, this model allows us to study how microscopic rules of individual behavior give rise to macroscopic regularities and global optimizations \cite{Epstein1999}.
The interaction substrate was defined by a random graph topology with a fixed average degree $\delta$ ranging between 4 and 6. In these experiments, we considered the colors $k$, $k=3,\dots,5$, to represent individual choices. Moreover, the single node payoff was represented by equation \eqref{node_payoff}, while social welfare by equation \eqref{social}. 

\begin{figure}[ht!]
\begin{tcolorbox}[colback=white, colframe=black!70, title=Box 2: Emergent Efficiency and the Saturation of Symbiosis, fonttitle=\bfseries]
    \centering
    \begin{tcolorbox}[colback=gray!5, colframe=white, size=small]
        \small
        {\bf Definition of Gain.} The Gain is the difference between the social welfare before and after the coalition has changed the coloring. 
        \begin{equation}
            Gain = \Delta \mu = \mu(\mathcal{V}, \gamma) - \mu(\mathcal{V}, \sigma)
        \end{equation}
        where $\sigma$ and $\gamma$ denote the system coloring before and after the coalition move.
        As shown in Figure \ref{fig:abm_results_box_A}, it measures the coordinated movements (symbiosis) that consistently reduce the percentage of unhappy nodes compared to purely individual dynamics. Gain in social welfare increases rapidly for small groups but exhibit a saturation once coalitions exceeds some threshold. Larger coalitions reach a lower state of conflict faster than smaller functional guilds.
    \end{tcolorbox}
\end{tcolorbox}
\end{figure}

\begin{figure}[!ht]
\centering
\includegraphics[width=0.8\textwidth]{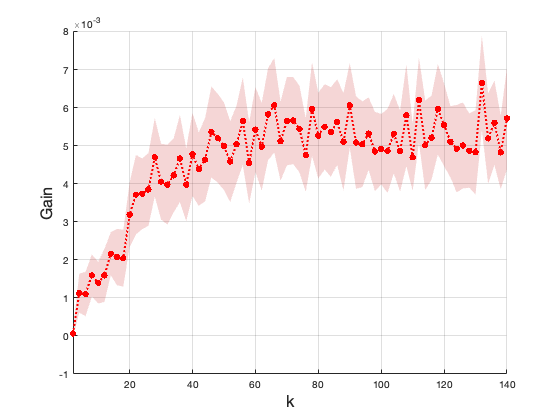}
\caption{{\bf Global Welfare Gain.}  The figure shows the gain, as defined in Box 2, of coalitions formed in the agent-based model. The positive values of the gain for all coalition sizes $k$ empirically validate the monotonic increase in social welfare during coordinated deviations. A saturation effect is observed for coalitions of more than 50 nodes.}
\label{fig:abm_results_box_A}
\end{figure}

The simulation followed a multi-stage evolutionary process: starting from a stochastic initial coloring, agents first pursue individual payoff maximization through decentralized best-response dynamics. Once the system has reached a Nash Equilibrium  -- where no single agent can unilaterally improve its fitness -- we allow the formation of coordinated deviations by symbiotic coalitions of different sizes $r$, let the nodes again free of changing color and eventually reach a new NE. The impact of coalitions on the system is evaluated by means of the gain defined in Box 2.

\begin{figure}
\centering
\includegraphics[width=0.8\textwidth]{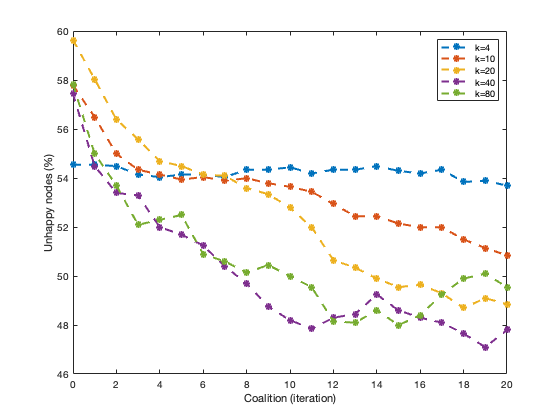}
\caption{{\bf Decay of Social Frustration.} The temporal evolution of unhappy nodes (agents in conflict) shows that iterations of coalition-forming drive the system toward higher payoffs. Notably, larger coalitions reach the metastable state faster, though the gain eventually plateaus as the network achieves an optimally partitioned SE.}
\label{fig:abm_results_box_B}
\end{figure}

Our results demonstrate a distinct transition in system efficiency. While purely individual dynamics often stall in sub-optimal local plateaus, the formation of functional guilds acts as a directional catalyst for global optimization, as it is shown by the gain evaluated in Figure \ref{fig:abm_results_box_A}. Figure \ref{fig:abm_results_box_B} illustrates the monotonic reduction of nodes that experience the stress test of being connected with nodes of the same color, with the social welfare gain increasing rapidly as the size of the coalition $r$ increases, eventually saturating for groups exceeding approximately 50 members. These simulations empirically confirm that coordination is essential to overcome topological frustration, leading the network toward a sequence of metastable biological states leading to a final climax, where the global social welfare is maximized, and exhaustive stress-tests confirm the absence of further profitable deviations. 

The simulation experiment provides empirical evidence for the saturation effect of symbiosis, highlighting how the collective drive toward efficiency eventually reaches a structural plateau consistent with our theoretical results proved by theorem \ref{thm:optimality} and conjecture \ref{thm:monotone}.

\subsection{Biological Application: From Taxonomic Chaos to Functional Climax}
\label{bio}

In this section we analyze the stability in ecological substrates following a comprehensive systematization, highlighting the intricate interplay between network architecture and community persistence \cite{Chen2024}. To demonstrate the ecological relevance of our theoretical framework, we applied the max $k$-cut game to a real-world pollination network from the Central Andes of Chile (Network ID 3, derived from \cite{THEBAULT2020, ARROYO1985}). This ecosystem presents an ideal stress test for our model: characterized by energy-limited alpine conditions and high resource saturation, nearly half of the plant species (48.8\%) are shared among multiple pollinators, creating an environment of intense competitive pressure. In such systems, stable coexistence is often achieved through the emergence of social dominance 
networks that partition resources to avoid direct agonistic encounters \cite{Fausch2021}.
In this context, we modeled 28 pollinator species as selfish agents, where an edge exists between two agents if they compete for the same floral resources.
\begin{figure}[!ht]
    \centering
    \begin{minipage}{0.45\textwidth}
        \centering    \includegraphics[width=\textwidth]{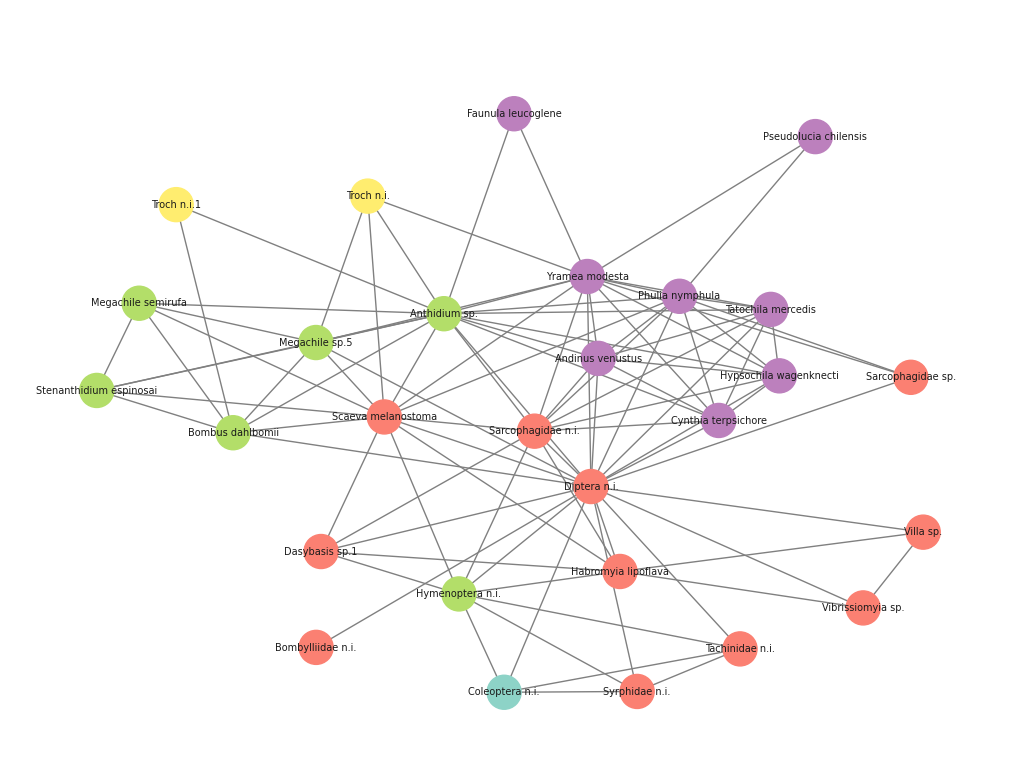}
        \small \textbf{(A) State 0: Taxonomic Chaos} \\ \scriptsize Payoff: 100
    \end{minipage}
    \hfill
    \begin{minipage}{0.45\textwidth}
        \centering     \includegraphics[width=\textwidth]{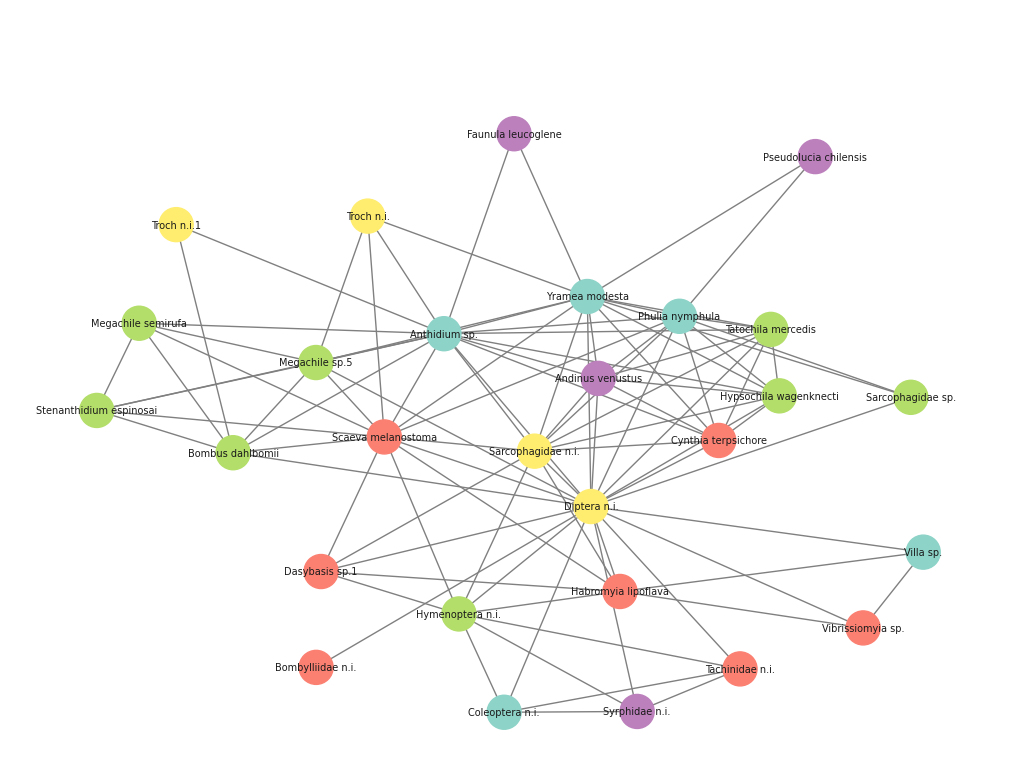}
        \small \textbf{(B) Step 3: Symbiotic Push} \\ \scriptsize Payoff: 166
    \end{minipage}


    \begin{minipage}{0.45\textwidth}
        \centering
        \includegraphics[width=\textwidth]{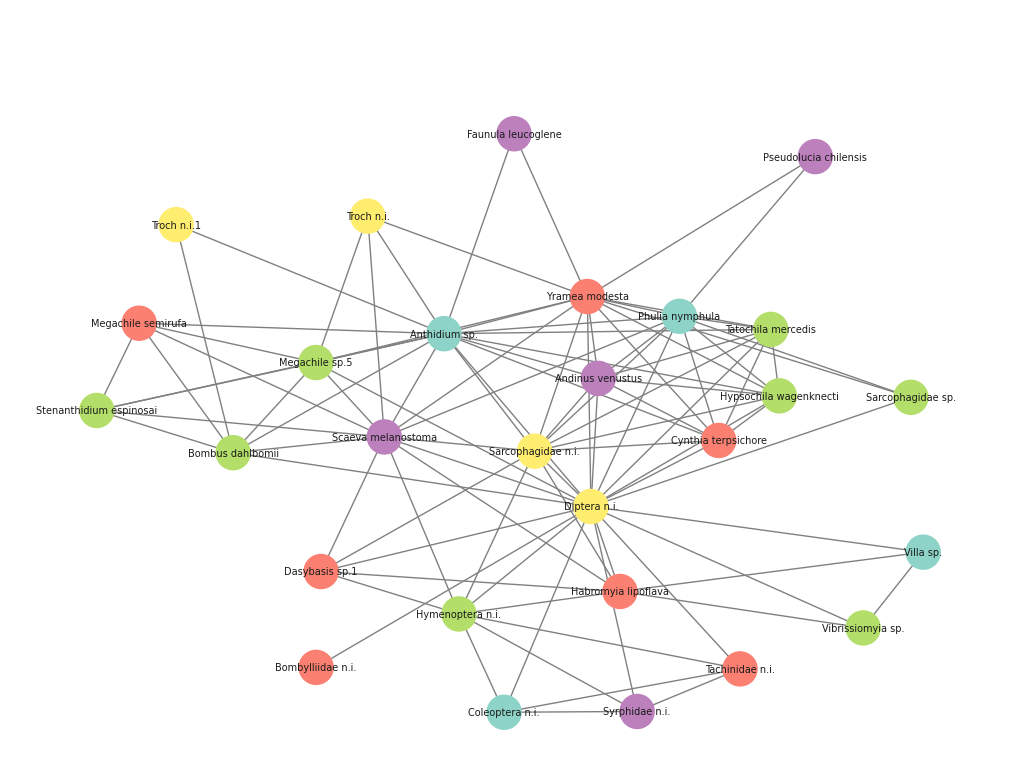}
        \small \textbf{(C) Step 4: Niche Refinement} \\ \scriptsize Payoff: 180
    \end{minipage}
    \hfill
    \begin{minipage}{0.45\textwidth}
        \centering    \includegraphics[width=\textwidth]{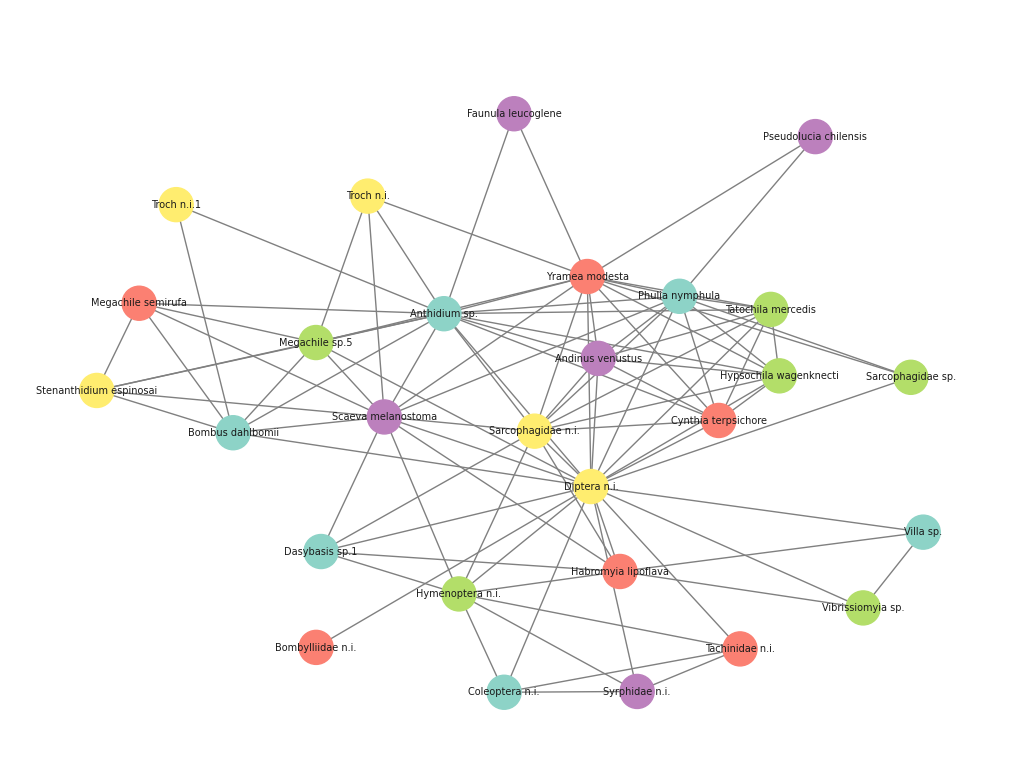}
        \small \textbf{(D) Final State: Biological Climax} \\ \scriptsize Payoff: 186
    \end{minipage}
\caption{Chronological evolution of the Network ID 3 pollination substrate. \textbf{(A. Initial)} The ancestral state is dominated by niche conservatism, where phylogenetic proximity leads to unsustainable resource competition (frustration). \textbf{(B \& C. Transition)} As coordinated guilds -- comprising both generalist and specialist species -- negotiate new traits, the system undergoes a series of strong deviations that increase global efficiency. This phase represents the active role of symbiosis as an optimization catalyst. \textbf{(D. Climax)} The final Climax state achieves maximum social welfare. 
}
\label{fig:evolution_climax_box}
\end{figure}

Our simulation traces the evolutionary trajectory of the system through three distinct phases. Initially, the network is set in an ancestral Taxonomic State, where species are grouped by phylogenetic proximity (e.g., Hymenoptera with Hymenoptera). This configuration reflects the principle of \textit{niche conservatism}, resulting in maximum structural frustration and a low global fitness (payoff = 100). Biologically, this represents a highly unstable state where taxonomic similarity leads to redundant resource use and excessive intraspecific competition.

The transition toward efficiency is driven by the emergence of Functional Guilds. Through a sequence of five coordinated symbiotic deviations (guilds of size $r=3$ or $4$), the pollinators negotiate new ecological niches. We observed that key generalist species, such as \textit{Diptera n.i.}, act as evolutionary motors; their strategic niche shifts trigger a drag effect (systemic altruism), passively improving the fitness of neighboring species by liberating previously contested resources.

\begin{figure}[hb!]
    \centering
\includegraphics[width=0.85\textwidth]{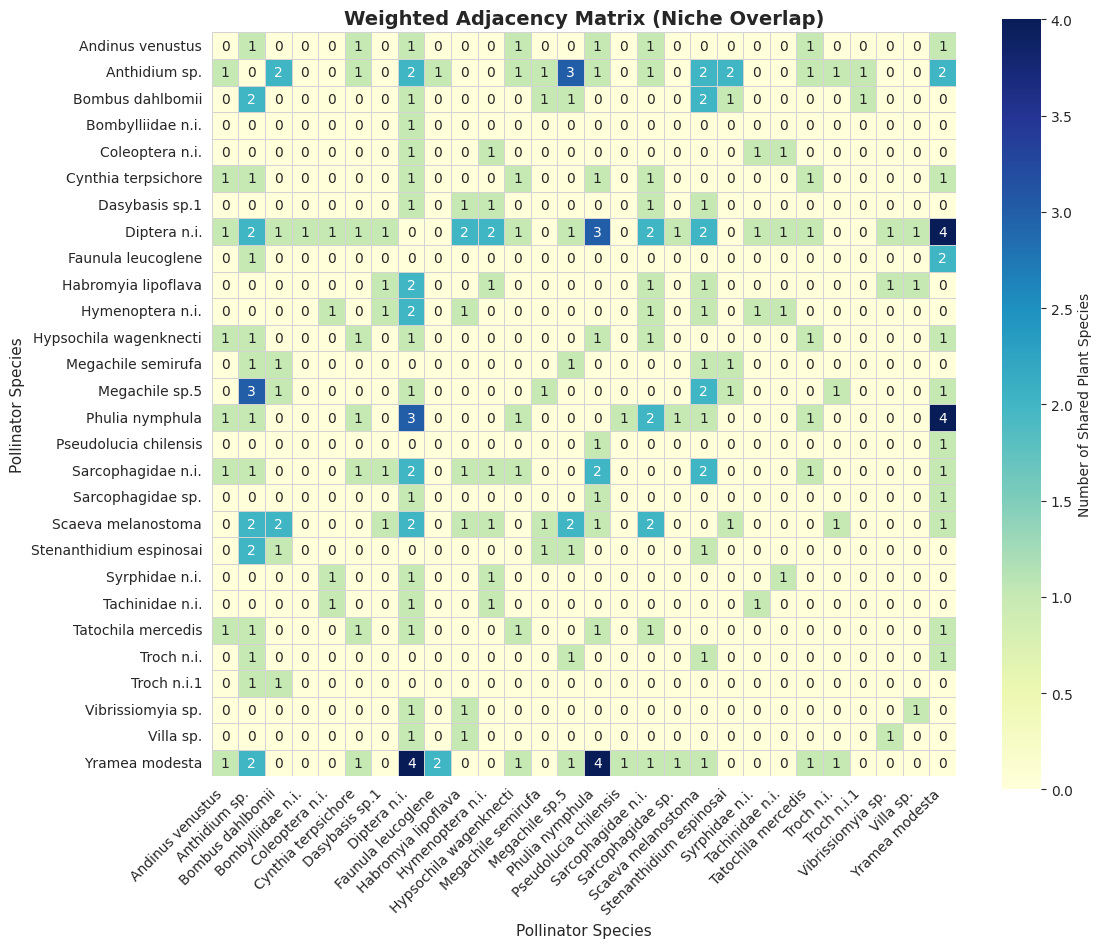}
\caption{\textbf{Weighted Adjacency Matrix representing niche overlap intensity.} In this refined representation of the Network ID 3, the edges are no longer binary. Instead, each cell $(i, j)$ represents the number of shared plant species visited by both pollinator $i$ and pollinator $j$, acting as a proxy for the \textit{strength} of interspecific competition. For instance, species such as \textit{Diptera n.i.} and \textit{Yramea modesta} exhibit high-intensity overlap (darker blue cells), indicating critical nodes where coordinated niche partitioning would yield the highest systemic benefit. }
\label{fig:weighted_matrix}
\end{figure}

The process culminates in a Biological Climax at a global optimum (payoff = 186), which we identify as a Strong Equilibrium. At this stage, the niche partitioning is so refined that no further coordinated shift can improve the fitness of any group. Interestingly, the final state is not a perfect zero-conflict landscape but a state of \textit{metastability}: five residual topological welding points (permanent conflicts) remain. These conflicts are not errors of the system but structural necessities; our model reveals that resolving these final tensions would cause negative externalities, undermining the global stability of the community. This highlights the crucial distinction between a static optimum and a dynamic ecosystem. The five residual 'welding points' we observe are not failures of the evolutionary process, but the very anchors that prevent the network from collapsing into a rigid, non-adaptive state. They ensure that the pollination network remains in a metastable regime, capable of further niche negotiations rather than being permanently locked in a structural sink.
This application confirms that the architecture of biodiversity is governed by laws of collective optimization, where symbiosis acts as the catalyst to reach a resilient, functionally integrated climax. This state of stability provides a game-theoretic foundation for the Red King effect, where symbiotic interactions -- such as those observed in lichens -- lead to a significant slowing of evolutionary rates as the system reaches a robust, optimal equilibrium \cite{Ametrano2021}.

The step-by-step transition from the initial taxonomic disorder to the optimized functional climax is visually detailed in Figure \ref{fig:evolution_climax_box}, which maps the topological reconfiguration of the pollinator network as functional guilds negotiate their respective ecological niches.

While our current model effectively identifies the structural stability of the ecosystem using an unweighted substrate, a natural extension for future research involves incorporating the \textit{intensity} of niche overlap. As illustrated in the Weighted Adjacency Matrix shown in Figure \ref{fig:weighted_matrix}, the competitive pressure between pollinators is not uniform but depends on the exact number of shared floral resources. Transitioning to a weighted max k-cut game will allow for a more granular representation of ecological tension, although it poses new theoretical challenges regarding the guaranteed existence of Strong Equilibria in non-discrete payoff landscapes. 

Considering weighted interaction substrate in future iterations of the max $k$-cut game will provide a deeper understanding of how the magnitude of resource competition influences the formation and stability of symbiotic functional guilds.

\subsection{Evolutionary Anthropology and the structural foundations of shared intentionality} \label{tomasello}

Beyond ecological niche partitioning, the strategic dynamics of the max $k$-cut game offer a compelling structural model for the major transitions in human social evolution. Our findings provide a mathematical foundation for the theories of Michael Tomasello regarding the emergence of \textit{Shared Intentionality} \cite{Tomasello2014, Tomasello2016, Tomasello1999, Tomasello2009}. In this framework, the transition from the individualistic, competitive strategies typical of great apes to the collaborative agency of early humans can be interpreted as a systemic escape from sub-optimal Nash Equilibria.

Our model demonstrates that in networks driven by purely individualistic incentives, the system inevitably becomes trapped in stagnant, low-welfare configurations (Nash traps). In these states, no single agent can improve its fitness alone, mirroring the Stag-Hunt dilemma where individualistic strategies prevent the achievement of superior collective goals. We show that the formation of a symbiotic coalition -- a coordinated shift in traits or strategies -- is not merely an option but a structural necessity to break this impasse. This transition is the exact topological analog of Tomasello?s \textit{Joint Agency}: the moment when individuals move beyond using others as social tools and begin to operate as a We, pursuing a shared goal that benefits the entire group.

Crucially, Conjecture \ref{thm:monotone} and its computational proof provide a formal basis for the ratchet effect in cultural evolution. By proving that symbiotic deviations strictly increase social welfare and generate positive externalities for non-participating members (Systemic Altruism), we show how the emergence of coordinated guilds creates a directional engine for community -- wide efficiency. In the human lineage, this structural pressure likely favored the evolution of cognitive architectures capable of managing dual-level intentionality -- where individuals maintain their own roles while simultaneously committing to a collective identity.

Finally, our proof that the global optimum is a Strong Equilibrium explains the long-term resilience of mature cooperative cultures. Once a society reaches an optimized state of mutualistic partitioning, the network architecture itself acts as a topological barrier against disruptive opportunistic mutations. Thus, the Biological Climax we identify represents the point where strategic altruism becomes locked into a stable social fabric, rendering the community resilient to the very individualistic greed that once characterized its ancestral state.

\section{Discussion}

The findings presented in this work provide a rigorous resolution to the long-standing diversity-stability debate, which has central importance in ecological theory \cite{McCann2000}, as well as to the conjecture regarding the resilience of optimal states in anti-coordination games. By proving that every global optimum in the max k-cut game is a Strong Equilibrium, we establish a fundamental principle of stability: in a system that has achieved maximum social welfare, collective greed is structurally inhibited. This perspective provides a formal mathematical basis for viewing optimized ecological communities as coherent interactors in the evolutionary process \cite{Lean2022}, where the community as a whole acts as a unified functional unit resilient to internal disruption. This results in what 
we term topological resilience, where the very architecture of the interaction network acts as a barrier against opportunistic mutations or coordinated deviations. Furthermore, our result provides a theoretical explanation for the observed mismatch between simulated extinction cascades and empirical observations \cite{Morton2022}, identifying the specific topological barriers that inhibit coextinction cascades and stabilize mature biodiversity. This concept of topological resilience complements the theory of alternative stable states, where a 
loss of resilience typically precedes sudden, catastrophic shifts in ecosystem structure \cite{Scheffer2001}.
In the context of evolutionary biology, our results offer a formal justification for the stability of the Biological Climax. We demonstrate a functional duality in symbiotic behavior: while the formation of coalitions serves as a powerful catalyst to overcome local sub-optimality and frustration in evolving populations, this drive becomes mathematically redundant once the ecosystem reaches its optimal niche partitioning. 

It is relevant to noticing that we assume  the system moves his evolution from a sub-optimal state, i.e. a Nash Equilibrium. Nevertheless, in such configurations, individual deviations are no longer profitable, yet coordinated deviations by groups of agents remain possible and can be effectively exploited to increase their collective payoff. We provide the first formal evidence that under specific hypotheses on colors disjunction and the computational evidence that, in the general case,  coalitions act as symbiotic drivers of systemic improvement, steering the network toward configurations of higher social welfare. We formalize a fundamental bridge between individual utilities and group-level topological properties, explicitly accounting for the reciprocal nature of interactions within a coalition. 

The described dynamics provide a structural foundation for the emergence of \textit{Shared Intentionality} in human evolution \cite{Tomasello2014}. Our model suggests that the cognitive shift from individual to joint agency was not a mere biological coincidence, but a necessary systemic response to escape the stagnant Nash traps of purely individualistic competition. By enabling the ratchet effect of cumulative cultural efficiency \cite{Tomasello1999}, coordinated coalitions allowed early human groups to reach higher states of social welfare that were previously inaccessible through isolated strategic moves.
Furthermore, the discovery of systemic altruism -- where the strategic shift of a symbiotic guild passively improves the fitness of non-participating inertial species -- suggests that evolution toward efficiency is a collective process that generates positive externalities across the entire network.

The discovery of this topological resilience suggests a new way to interpret biodiversity. If the global optimum is a static sink, then the persistent frustration observed in nature is not a failure of optimization, but a functional necessity. The succession of symbioses we observe is an adaptive tracking process: the system is trapped in a productive metastable state, where coalitions are constantly negotiating the next incremental improvement, avoiding the terminal stasis of the global peak.

Beyond theoretical biology, these results have significant implications for the design of Artificial Intelligence and Multi-Agent Systems. In distributed task allocation, where agents must minimize interference (represented in our model by shared colors), our proof guarantees that a socially optimal allocation is robust not only to individual failures but also to adversarial manipulations by coordinated clusters of agents. Moreover, the algebraic decomposition of payoffs derived here could provide a natural potential function for multi-agent reinforcement learning algorithms, enabling agents to learn strategic altruism by aligning local updates with global stability.

In the realm of computational sociology, our framework provides insights into collective cohesion and the resilience to polarization. If a social system is configured to reward anti-coordination -- promoting interaction with diverse viewpoints -- our findings suggest that such a system is inherently protected from the formation of isolated echo chambers or polarized coalitions. Coordinated groups cannot deviate to increase their own isolation without suffering a loss in individual utility, provided the global state is sufficiently optimized for diversity.

Future research should focus on extending this proof to weighted and directed graphs, where, as noted in the literature, the discreteness of individual improvements is lost and the existence of Strong Equilibria is not always guaranteed. Additionally, applying this max $k$-cut lens to large-scale empirical datasets, such as the pollination networks analyzed in recent eco-evolutionary studies, could unveil the specific welding points of competition that maintain the delicate balance of global biodiversity.

We notice that our findings carry profound implications that extend beyond the immediate scope of graph theory and algorithmic game theory. The structural inability of the network to lock into a static global optimum -- driven by the perpetual emergence and dissolution of parasitic coalitions -- suggests that these systems naturally operate in a metastable regime. Rather than viewing this lack of convergence as a computational failure, it should be recognized as an intrinsic signature of complex adaptive networks. This theoretical framework resonates deeply with empirical phenomena observed across diverse disciplines: from the transient attractors governing neural dynamics in the human brain, to the shifting alliances in political ecosystems, and the continuous adaptive tracking required for socio-ecological sustainability. By demonstrating that perpetual reconfiguration is a fundamental feature rather than a bug, our model provides a robust mathematical foundation for understanding metastability not just as a transitional phase, but as the true engine of long-term resilience in evolving systems.

Moreover, our work suggests that 'optimal' and 'stable' are not synonymous with 'static.' By proving that coalitions are impossible only in the global optimum, we identify a fundamental boundary: the network architecture itself acts as a stabilizer that protects the system's reached efficiency, while the relentless cycle of symbiotic formation and dissolution provides the necessary fuel for continuous adaptation. In the balance between the stasis of the optimum and the flux of metastability, we find the true engine of long-term resilience.

The interpretation of shared intentionality  \cite{Tomasello2014, Tomasello2016, Tomasello1999, Tomasello2009} as a catalyst for systemic efficiency proposed in our paper, finds a crucial technical parallel in the recent findings published in \cite{LeCun}, where the authors argue that the current inability of artificial intelligence to learn flexibly stems from the lack of an architecture that integrates social signals and communication --functions that are presently 'outsourced' to human experts. Mirroring our model, where the formation of coalitions allows agents to escape sub-optimal Nash traps, a System M (Meta-control) is identified as the architectural tool required to prioritize social cues and goal imitation. This structural flexibility is deeply rooted in the concept of metastability, a regime where a system coexists between integration and segregation, avoiding becoming a 'prisoner' to a single, frozen state. In this context, the emergence of symbiotic coalitions in our max k-cut game represents a metastable dynamic where productive frustration is not a failure of optimization, but a functional necessity for continuous adaptation. This transition serves as the topological foundation for the ratchet effect cited in \cite{LeCun}, ensuring that gains in collective welfare and knowledge are protected by topological barriers against individualistic regression while maintaining the system's vital capacity for perpetual reconfiguration.

Ultimately, this theoretical framework of metastability extends far beyond algorithmic game theory, offering a rigorous mathematical lens to decode some of the most pressing contemporary global challenges. For instance, in the context of global climate action, international agreements, such as the UN Climate Change Conferences, often struggle to maintain long-term stability despite the clear, collective theoretical optimality of global cooperation. Our model provides a structural explanation for this paradox: such political or environmental alliances do not simply 'fail' due to irrationality or lack of commitment. Rather, they are subject to the same divergent 'parasitic' dynamics and cross-scale conflicts that drive our unweighted networks into continuous reconfiguration. Similarly, in financial markets, the perpetual formation and dissolution of strategic alliances -- such as ethical divestment coalitions -- reflect a system living and transitioning through metastable states, constantly adapting to shifting micro-incentives. By framing these phenomena not as systemic anomalies or policy failures, but as natural structural expressions of metastable resilience, our findings suggest a crucial paradigm shift: effective global policies must be designed to manage and navigate continuous adaptation, rather than relying on the fragile illusion of static consensus.

\clearpage 





\end{document}


\maketitle

\section{Introduction}
The study of complex networks has traditionally focused on individual stability, often modeled through the lens of Nash Equilibria, \cite{Nash1951}. However, in biological ecosystems, species do not always act in isolation; they form functional guilds or symbiotic coalitions to overcome environmental constraints \cite{Odum1969}. A fundamental question arises: can a group of opportunistic agents, acting in their own collective interest, destabilize a state that is already socially optimal? 

In this work, we model ecological niche partitioning as a strategic max $k$-cut game \cite{Panagopoulou2008,Gourves2009,Escoffier2010}. We prove that in unweighted interaction networks, any state that maximizes the global welfare (the social optimum) is inherently immune to any form of coalitional deviation \cite{Gourves2010}. This result provides a mathematical foundation for the concept of biological climax as a state of maximum systemic resilience \cite{Holling1973}.

\section{Preliminaries and Problem Formulation}\label{preliminaries}

In this section, we formally introduce the mathematical framework of the strategic max $k$-cut game played over a graph. The {\em cut} of a graph is the set of edges that connect nodes of different colors. The notation follows the standard conventions of graph theory and non-cooperative game theory, specifically focusing on the tension between individual incentives and global social welfare. 

The interaction substrate is modeled as an undirected, unweighted graph $G = (\V, \E)$, where $\V = \{1, \dots, N\}$ represents the set of \textit{nodes} (or players), with $N \geq 2$ and $\E \subseteq \V \times \V$ is the set of \textit{edges} representing mutual interactions or constraints between players.
We assume the graph is simple, meaning it is devoid of self-loops ($a_{v,v} = 0$) and multiple edges. The structure of $G$ is fully described by its symmetric adjacency matrix $A = \{a_{v,w}\}_{v,w \in \V}$, where $a_{v,w} = 1$ if $\{v, w\} \in \E$, and $0$ otherwise.
For each node $v \in \V$, the neighborhood of $v$ is the set $N(v) = \{w \in \V : a_{v,w} = 1\}$, and the degree of node $v$ is defined as $\delta_v = \sum_{w \in \V} a_{v,w}$, which represents the total number of neighbors of player $v$.

In a max $k$-cut game, each player $v \in \V$ must choose a strategy from a finite set of $k \geq 2$ available colors, denoted by $\K = \{1, 2, \dots, k\}$. The following definition introduces the concept of the coloring of a graph.

\begin{definition}[Coloring / Strategy Profile]
A coloring (or strategy profile) is a vector $\sigma = (\sigma_1, \sigma_2, \dots, \sigma_N) \in \K^N$, where $\sigma_v \in \K$ denotes the color chosen by player $v$.
\end{definition}

The set $\K^N$ represents the state space of the system, containing all possible assignments of colors to nodes.


The objective of each player is to maximize their individual \textit{fitness} through anti-coordination. This reflects scenarios where players seek to differentiate themselves from their neighbors to avoid interference or resource competition. 


\begin{definition}[Individual Payoff and Conflict Degree]\label{individual_payoff} 
The payoff of player $v$ under the coloring $\sigma$ is defined as the number of neighbors who have chosen a color different from their own:
\begin{equation}
\mu_v(\sigma) = \sum_{w \in \V} a_{v,w} \mathbbm{1}(\sigma_w \neq \sigma_v) \label{ind_payoff}
\end{equation}
where $\mathbbm{1}(\cdot)$ denotes the indicator function, which equals $1$ if the condition inside the parentheses is satisfied, and $0$ otherwise. Conversely, we define the conflict degree of player $v$ as the number of neighbors sharing its exact same color:
\begin{equation}
c_v(\sigma) = \sum_{w \in \V} a_{v,w} \mathbbm{1}(\sigma_w = \sigma_v).
\end{equation}
Finally, we define the color degree of a player $v$ as the number of their neighbors of color $k$:
\begin{equation}
n_v(k,\sigma) = \sum_{w \in \V} a_{v,w} \mathbbm{1}(\sigma_w = k),
\end{equation}
\end{definition}

Note that $\mu_v(\sigma) \leq \delta_v$. Moreover, since the underlying topology of the network is fixed, the sum of a node's payoff and its conflicts is inherently conserved and equals its total degree: $\mu_v(\sigma) + c_v(\sigma) = \delta_v$. If $\mu_v(\sigma) = \delta_v$, the node is locally satisfied with respect to all its edges, meaning $c_v(\sigma) = 0$. We additionally note that $n_v(\sigma_v,\sigma) = c_v(\sigma)$.

As a direct consequence of the topological conservation $\mu_v(\sigma) + c_v(\sigma) = \delta_v$, any strategy variation that increases a player's individual payoff implies a strict, proportional decrease in their total conflicts. Formally, for any two configurations $\sigma$ and $\gamma$, the variation in payoff is exactly the opposite of the variation in conflicts:
\begin{equation}
\Delta \mu_v(\gamma, \sigma) = - \Delta c_v(\gamma, \sigma).
\end{equation}
Therefore, a player performs a strict improvement ($\Delta \mu_v > 0$) if and only if their total number of conflicts strictly decreases ($\Delta c_v \le -1$).

From a system-wide perspective, we are interested in the {\em global payoff}, or {\em social welfare}, which corresponds to the total utility of the population. In the following, we present the definitions and preliminary concepts that lead to our results. 

\begin{definition}[Global Payoff, Social Welfare and Cut Value]
The global payoff, or social welfare, $\mu(\V, \sigma)$ is the sum of all individual payoffs:
\begin{equation}
    \mu(\V, \sigma) = \sum_{v \in \V} \mu_v(\sigma).\label{global_payoff}
\end{equation}
Moreover, the global conflict, or social unhappiness, is the sum of all individual conflicts:
\begin{equation}
    c(\V, \sigma) = \sum_{v \in \V} c_v(\sigma).\label{global_coflict}
\end{equation}
\end{definition}
In combinatorial terms, $\mu(\V, \sigma)$ is exactly twice the value of the cut induced by the partition of colors, as each cut edge is counted once for each of its two endpoints. The same holds for the conflicts indicator $c(\V,\sigma)$, which counts each conflicting edge twice. 

The above definitions can be used to calculate the payoff and the conflict of a given set $F \subset \V$, thus $\mu(F,\sigma)=\sum_{v\in F} \mu_v(\sigma)$ and $c(F,\sigma)=\sum_{v\in F} c_v(\sigma)$, where $\mu_v(F,\sigma)$ and $c_v(F,\sigma)$ are the payoff and the conflict of the nodes in the subset $F$, respectively.

We can consider also the cut between two different sets. Let $S_1, S_2 \subseteq \V$ be two subsets of nodes (not necessarily disjoint). We define the set of edges connecting $S_1$ to $S_2$ as $\E_{s}(S_1, S_2) = \{\{v, w\} \in \E : v \in S_1, w \in S_2\}$.

\begin{definition}[Cut between Sets]\label{cut_sets}
Given a coloring $\sigma$, we define $P(S_1, S_2, \sigma)$ as the number of cut edges, i.e. \ edges with endpoints of different colors, in the set $\E_s(S_1, S_2)$:
\begin{equation}
P(S_1, S_2, \sigma) = \sum_{v \in S_1} \sum_{\substack{w \in S_2 \\  \sigma_w \neq \sigma_v}} \frac{a_{v,w}}{\eta},\label{cut_between_eq}
\end{equation}
where $\eta = 2$ if the edge is counted twice, i.e. when $v, w$ are in the same set, and $\eta = 1$ otherwise. \label{cut_between}
\end{definition}

 One of the central concepts of this work is the {\em global optimum}, related to the configuration of maximum collective efficiency. The global optimum is also called {\em social optimum}.

\begin{definition}[Global Optimum]
A coloring $\sigma^*$ is a global optimum if it maximizes the global payoff:
\begin{equation}
    \sigma^* \in \arg \max_{\sigma \in \K^N} \mu(\V, \sigma).
\end{equation}
\end{definition}

To analyze the system's effectiveness in finding and maintaining an optimal state, we evaluate how groups of players might act individually or coordinate to improve their payoff. To this aim, we define {\em deviations} and {\em strong deviations}, where, in the latter, every member of the group must strictly benefit from the change.

\begin{definition}[Deviating Coalition]
Given a coloring $\sigma$, a subset of nodes $C \subseteq \V$ is a deviating coalition if there exists a new coloring $\gamma$ such that:
\begin{itemize}
    \item $\gamma_v \neq \sigma_v$ for all members of the coalition ($v \in C$).
    \item $\gamma_v = \sigma_v$ for all players outside the coalition ($v \notin C$).
\end{itemize}
\label{coaldef}
\end{definition}

\begin{definition}[Strong Deviation]
\label{devstrong}
A coalition $C$ performs a strong deviation from $\sigma$ to $\gamma$ if all its members strictly improve their individual payoff:
\begin{equation}
    \Delta \mu_v(\gamma, \sigma) = \mu_v(\gamma) - \mu_v(\sigma) > 0, \quad \forall v \in C.\label{strong}
\end{equation}
\end{definition}

Moreover, we introduce the concepts of  {\em Nash equilibrium} (NE), for which no single individual can improve their payoff by unilaterally changing their color, and {\em strong Nash equilibrium}, preventing groups of individuals from performing any profitable coordinated color changes. 

\begin{definition}[Equilibria]
A coloring $\sigma$ is a:
\begin{itemize}
    \item Nash Equilibrium (NE) if no coalition of size one (i.e., no individual player) can perform a strong deviation. The Nash equilibrium corresponds to a sub-optimal configuration.
    \item Strong Equilibrium (SE) if no coalition $C \subseteq \V$ of \textit{any} size can perform a strong deviation.
\end{itemize}
\end{definition}

We introduce a particular kind of coalition for which the set of colors they can choose is restricted to those not present in the boundary set $D$.

\begin{definition}
    Let $F$ be a deviating coalition and $D$ be its boundary. A deviation to a new configuration $\gamma$ is boundary-disjoint if the set of strategies (or colors) adopted by the nodes in $F$ under $\gamma$ does not intersect with the set of strategies used by the nodes in $D$. Formally: 
    \begin{equation}
   \mathcal{S}(F_\gamma) \cap \mathcal{S}(D) = \emptyset. \label{disjoint}
   \end{equation}
\end{definition}

Finally, we introduce metastable configurations as intermediate configurations that are stable against individual moves but vulnerable to collective ones. 

\begin{definition}[Metastable Configuration and Resilience Rank]
A coloring $\sigma$ is said to be metastable if it is a NE but not a SE. Formally:
\begin{equation}
\sigma \in \{ \sigma \in \mathcal{K}^N \mid \forall v \in \mathcal{V}, \Delta \mu_v \le 0 \} \setminus \{ \sigma \in \mathcal{K}^N \mid \text{no coalition } F \text{ can perform a strong deviation} \}.\label{metastable}
\end{equation}
Furthermore, the stability of such states is quantified by their resistance to collective perturbations. A metastable coloring $\sigma$ is defined as \textbf{$m$-rank resilient} if no coalition $F \subseteq \mathcal{V}$ with size $|F| < m$ can perform a strong deviation.
\end{definition}



The central finding of this work is to demonstrate that in unweighted max $k$-cut games, every global optimum is not only a Nash Equilibrium but also a Strong Equilibrium, ensuring total resilience against collective opportunistic mutations.

\section{Preliminary Results}\label{preliminary_res}

In this section, we establish the formal link between the welfare of individual members of a coalition and the global payoff of the system. 

Given a coalition $F \subseteq \V$ we can identify the following subsets of $\mathcal{V}$.

\begin{itemize}
    \item $C(\sigma)$ is the set of \textit{unsatisfied nodes}, i.e. those who could potentially improve their payoff. The coalition $F$ is a subset of $C$.
    \item $D = C(\sigma) \setminus F$ is the set of \textit{inertial candidates}, i.e. nodes that could move but stay still.
    \item $E = V \setminus (F \cup B)$ is the set of \textit{external nodes}, i.e. the ones not connected to the coalition.
\end{itemize}


 In the rest of our paper, we assume the population has already reached a suboptimal configuration, i.e. it is in a Nash equilibrium, so that it can be effectively able to profit from groups coalition thus allowing them to increase their own payoff and driving the whole system towards an improved state. In the following Lemma, we establish a fundamental bridge between individual utilities and group-level topological features, accounting for the reciprocal nature of interactions within a coalition. This decomposition  is crucial for isolating the internal synergy of a group from its external interactions with the rest of the network.

\begin{lemma}[Decomposition of Coalitional Payoff]\label{decompose}
For any sub-optimal coloring $\sigma$ and any coalition $F \subseteq \V$, the following identity holds:
\begin{equation}
\sum_{v \in F} \mu_v(\sigma) = 2 \cdot P(F, F, \sigma) + P(F, \V \setminus F, \sigma).
\end{equation}
\label{deccoalpay}
\end{lemma}

\begin{proof}
From the definition of individual payoff in equation \eqref{ind_payoff}, and the cut between sets in equation \eqref{cut_between_eq}, we have that:
\begin{equation}
\sum_{v \in F} \mu_v(\sigma) = \sum_{v \in F} \left( \sum_{\substack{w \in F \\ \sigma_w \neq \sigma_v}} a_{v,w} + \sum_{\substack{w \in \V \setminus F\\ \sigma_w \neq \sigma_v}} a_{v,w} \right).
\end{equation}
The first term in parentheses counts cut edges whose endpoints are both in $F$. Since the graph is undirected, each edge $\{v, w\}$ is counted twice. Therefore, this term is equal to $2 \cdot P(F, F, \sigma)$.
 The second term in parentheses counts cut edges connecting a node in the coalition to an external node. Each such edge has only one endpoint in $F$, so it is counted exactly once, then it is $P(F, \V \setminus F, \sigma)$. Since all nodes have been considered, the proof is complete.
\end{proof}

 
The following identity establishes a formal link between micro-scale strategic moves and macro-scale systemic outcomes. It allows us to calculate the net variation in global social welfare by aggregating individual payoff improvements, while explicitly correcting for the reciprocal effects of internal coordination within the deviating group:

\begin{lemma}[Global Variation Identity]\label{variation_identity}
The variation of the global payoff produced by the deviation $F$ from the coloring $\sigma$ to the sub-optimal coloring $\gamma$ is related to the variation of individual payoffs by the following relation:
\begin{equation}
\Delta\mu(\V, \gamma, \sigma) = 2 \sum_{v \in F} \Delta\mu_v(\gamma, \sigma) - 2\Delta P(F, F),
\end{equation}
where $\Delta P(F, F) = P(F, F, \gamma) - P(F, F, \sigma)$ is the cut variation between two sets and $\Delta \mu(\mathcal{V},\gamma,\sigma) = \mu(\mathcal{V},\gamma) - \mu(\mathcal{V},\sigma)$ is the global payoff variation.
\end{lemma}

\begin{proof}
Recall that the global social welfare is exactly twice the value of the total cut of the graph: $\mu(\mathcal{V}, \sigma) = 2 \cdot P(\mathcal{V}, \mathcal{V}, \sigma)$. We can decompose the total cut by partitioning the edges into three categories: those internal to the coalition $F$, those connecting $F$ to the rest of the network $\mathcal{V} \setminus F$, and those internal to the remaining nodes:
\begin{equation}
P(\mathcal{V}, \mathcal{V}, \sigma) = P(F, F, \sigma) + P(F, \mathcal{V} \setminus F, \sigma) + P(\mathcal{V} \setminus F, \mathcal{V} \setminus F, \sigma).
\end{equation}
When the coalition $F$ deviates from $\sigma$ to $\gamma$, only the nodes in $F$ change their colors. Consequently, the cut status of edges not incident to $F$ remains unchanged, implying that the variation $\Delta P(\mathcal{V} \setminus F, \mathcal{V} \setminus F) = 0$. Then, by the definition of variation of cut between sets, we have that the global variation is:
\begin{equation}
\Delta\mu(\mathcal{V}, \gamma, \sigma) = 2 \left[\Delta P(F, F) + \Delta P(F, \mathcal{V} \setminus F)\right]. \label{eq:global_var_step}
\end{equation}

Now, considering the sum of individual payoffs variation for the members of $F$ and using Lemma \ref{deccoalpay}, we have:
\begin{equation}
\sum_{v \in F} \Delta\mu_v(\gamma, \sigma) = 2\Delta P(F, F) + \Delta P(F, \mathcal{V} \setminus F).
\end{equation}
From this expression, we can isolate the term representing the variation of the cut between the coalition and the external network:
\begin{equation}
\Delta P(F, \mathcal{V} \setminus F) = \sum_{v \in F} \Delta\mu_v(\gamma, \sigma) - 2\Delta P(F, F). \label{eq:substitution}
\end{equation}
Finally, by substituting equation \eqref{eq:substitution} into the expression for global variation \eqref{eq:global_var_step}, we obtain:
\begin{equation}
\Delta\mu(\mathcal{V}, \gamma, \sigma) = 2 \left[\Delta P(F, F) + \sum_{v \in F} \Delta\mu_v(\gamma, \sigma) - 2\Delta P(F, F)\right].
\end{equation}
Simplifying the terms inside the brackets leads to the final identity:
\begin{equation}\label{sedici}
\Delta\mu(\mathcal{V}, \gamma, \sigma) = 2 \sum_{v \in F} \Delta\mu_v(\gamma, \sigma) - 2\Delta P(F, F),
\end{equation}
which concludes the proof.
\end{proof}


In the following, we establish an upper bound on the ability of a coalition $F$ to increase the number of internal cut edges. First of all, we establish that a coalition $F$ satisfies the \textit{low-conflict density} condition if the number of internal conflicting edges is strictly less than the number of nodes. 

To evaluate the potential for internal improvement within a deviating group, we must characterize the pre-existing conflicts that the coalition aims to resolve. The following lemma establishes a fundamental topological ceiling: the growth of the internal cut is strictly limited by the number of initial same-colored edges within the group. 

\begin{lemma}[Internal Cut Growth Constraint]\label{lemma:conflicts_density}
Let $\sigma$ be a Nash Equilibrium and let $F \subseteq V$ be a deviating coalition. When the agents in $F$ perform a joint deviation to a new configuration $\gamma$, the net variation of the internal cut is strictly bounded by the size of the coalition. Formally:

\begin{equation}
\Delta P(F, F) < |F|.\label{low_density}
\end{equation}
\end{lemma}

\begin{proof}
Let $E_{same}(F, \sigma)$ be the set of internal conflicting (same-colored) edges in $F$ under $\sigma$. The variation of the internal cut is given by the difference between the resolved and the newly created internal conflicts:
\begin{equation}\Delta P(F,F) = |E_{same}(F, \sigma)| - |E_{same}(F, \gamma)| \le |E_{same}(F, \sigma)|. \label{<E_same}
\end{equation}
Assume by contradiction that $|E_{same}(F, \sigma)| \ge |F|$. By graph theory, this implies that the subgraph induced by $E_{same}(F, \sigma)$ contains at least one closed cycle $C \subseteq F$. Let $c_v(\sigma)$ be the total number of conflicts of a node $v$ in configuration $\sigma$. By definition of strong deviation, every node $v \in C$ must strictly decrease its total conflicts in the new configuration $\gamma$:
\begin{equation}c_v(\gamma) < c_v(\sigma) \quad \forall v \in C. \label{strong_variation}
\end{equation}

Since $\sigma$ is a Nash Equilibrium, node $v$ cannot improve its payoff by unilaterally switching to any other color $k \in \mathcal{K} \setminus \{\sigma_v\}$. Let $n_v(k, \sigma) = \sum_{w \in \mathcal{V}} a_{v,w} \mathbbm{1}(\sigma_w = k)$ be the number of neighbors of $v$ sharing color $k$. This means that $n_v(k, \sigma)$ cannot exceed the number of neighbors sharing $v$'s current color, which is $c_v(\sigma)$. Specifically:
\begin{equation}
n_v(k, \sigma) \leq c_v(\sigma) \quad \forall k \in \mathcal{K}.
\label{bounds}
\end{equation}

This implies that the external environment of $C$ is saturated. If the nodes in $C$ shift to a new color configuration $\gamma$, the sum of their new conflicts must account for both the external bounds as in equation \eqref{bounds} and the internal edges within the cycle. Because $C$ is a closed topological loop, any reassignment $\gamma$ inevitably forces at least one node $u \in C$ to either hit the external saturation bound or recreate an internal conflict within $C$, resulting in:
\begin{equation}
c_{u}(\gamma) \ge c_{u}(\sigma). \label{cvgamma_cvsigma}
\end{equation}
Equation \eqref{cvgamma_cvsigma} directly contradicts the strong deviation requirement in equation \eqref{strong_variation}. This structural impossibility proves that $F$ cannot contain any internal conflict cycles. Thus, the conflict subgraph must be a collection of acyclic components (a forest), which requires:\begin{equation}|E_{same}(F, \sigma)| \le |F| - 1 < |F|.\label{f-1}
\end{equation}
Substituting \eqref{f-1} into \eqref{<E_same}, we obtain the thesis:
\begin{equation}\label{soglia25}
\Delta P(F,F) < |F|.
\end{equation}

\end{proof}

A natural conjecture is that for every individual node $v \in F$ performing a strong deviation from a NE, the number of boundary conflicts, $c_v, \forall v \in D$, does not increase pointwise. However, this claim does not follow from the NE condition alone. The NE condition $n_v(k,\sigma) \leq c_v(\sigma), \forall k$ provides a bound on total neighbors sharing the target color across all nodes of $\V$, but cannot be restricted to $D$ without additional assumptions. If $v$ holds internal conflicts ($c_v(\sigma) > 0$), the bound permits that in the set $D$ an individual node may increase its boundary conflicts during the deviation ($n_v(\gamma_v, \gamma) > c_v(\sigma)$).

This motivates the complementary results of the following Section \ref{sec4}. Theorem \ref{bbeth} establishes a universal lower bound on the aggregate boundary variation, while Theorem \ref{bound_disj} establishes non-negativity under the boundary-disjoint condition. Together, these results bracket the aggregate behavior of the frontier $D$, and are supported by the computational validation extensively reported in the Main Document.

\section{Main Results}\label{sec4}

In this section, we first present the core result of this work: the formal proof that any coloring that maximizes the global payoff, or the social optimum, is intrinsically protected against collective deviations of any size.

\begin{theorem}[Optimality $\implies$ Strong Equilibrium]
Let $G = (\V, \E)$ be an unweighted and undirected graph. Every optimal coloring $\sigma^*$ for the max $k$-cut game is a Strong Equilibrium (SE).
\label{3punto1}
\end{theorem}

\begin{proof}
The proof proceeds by contradiction. Suppose that a globally optimal coloring $\sigma^*$ is \textit{not} a Strong Equilibrium. By the definition of a Strong Equilibrium, this implies the existence of at least one non-empty coalition $F \subseteq \mathcal{V}$ and a new coloring $\gamma$ capable of performing a \textit{strong deviation}. In such a scenario, every member $v \in F$ must strictly improve their individual utility ($\Delta\mu_v > 0$). Using the definition \ref{individual_payoff} of individual payoff,  we can calculate its variation as:
\begin{equation}\label{venti4}
\Delta\mu_v(\gamma, \sigma^*) = \mu_v(\gamma) - \mu_v(\sigma^*) \geq 1, \quad \forall v \in F,
\end{equation}
and by summing the individual improvements over the entire coalition, we obtain:
\begin{equation}
\sum_{v \in F} \Delta\mu_v(\gamma, \sigma^*) \geq |F|. \label{eq:sum_bound}
\end{equation}

To evaluate the systemic effect of this deviation, we invoke the Global Variation Identity proven in Lemma \ref{variation_identity}, which allows us to express the change in social welfare as:

By applying the \textit{Global Variation Identity} shown in Lemma \ref{variation_identity}, we can express the change of the social welfare as a function of individual incentives and the internal cut edges $\Delta P(F, F)$:
\begin{equation}
\Delta\mu(\mathcal{V}, \gamma, \sigma^*) = 2 \left( \sum_{v \in F} \Delta\mu_v(\gamma, \sigma^*) - \Delta P(F, F) \right). \label{eq:final_relation}
\end{equation}

 Plugging equation \eqref{eq:sum_bound} into equation \eqref{eq:final_relation} we have that: 
 \begin{equation}
\Delta\mu(\mathcal{V}, \gamma, \sigma^*) \geq 2(|F| - \Delta P(F, F)). \label{2f-DP}
\end{equation}

 Since any globally optimal configuration $\sigma^*$ maximizes the social welfare, no single node can strictly improve its payoff without decreasing the global sum; therefore, $\sigma^*$ is necessarily a Nash Equilibrium. This allows us to apply the Internal Cut Growth Constraint from Lemma \ref{lemma:conflicts_density}, which states that for any coalition deviating from a Nash Equilibrium, $\Delta P(F,F) < |F|$. 
 
 Substituting the strong deviation condition ($\sum_{v \in F} \Delta\mu_v \ge |F|$) and the topological bound from Lemma \ref{lemma:conflicts_density} into equation \eqref{2f-DP} we get:
\begin{equation}
\Delta\mu(\mathcal{V},\gamma,\sigma^*) > 2(|F| - |F|) = 0.
\end{equation}
 
 This strict inequality $\Delta\mu(\mathcal{V},\gamma,\sigma^*) > 0$ contradicts the hypothesis that $\sigma^*$ is a global optimum. Thus, no such strong deviation can exist, proving that $\sigma^*$ is a Strong Equilibrium.
\end{proof}

The above theorem proves that in a max $k$-cut game on unweighted graphs, no coalition $F$ can perform a strong deviation from an optimal coloring. Therefore, every global optimum $\sigma^*$ is necessarily a Strong Equilibrium. To prove it, the assumption of unweighted edges is crucial. In a weighted graph, a coalition might find a deviation where the sum of their individual gains outweighs the loss of a few high-value edges internally. In such cases, a global optimum might not be a Strong Equilibrium. The unitary weight of biological interactions in our model represents the discrete nature of niche overlap in saturated environments.\\

Now we investigate if a strong deviation, despite being driven by the purely selfish incentives of the members of a coalition $F$, can produce a systemic benefit. 

We assume that $\sigma$ is a sub-optimal coloring that admits a strong deviation towards $\gamma$ and consider the set $D = C(\sigma) \setminus F$ of inertial candidates who do not actively participate in the color change. We demonstrate that the payoff variation for these nodes is lower bounded.

\begin{theorem}[Bounded Boundary Externalities]\label{bbeth}
    Let $\sigma$ be a Nash Equilibrium and $F$ be a coalition performing a strong deviation to a new configuration $\gamma$. Let $D$ be the boundary of $F$ (i.e., the set of nodes strictly outside $F$ that share at least one edge with a node in $F$). The total payoff variation experienced by the boundary $D$ is strictly bounded from below by $2 - |F|$. Formally:$$\sum_{v \in D} \Delta\mu_v(\gamma, \sigma) \ge 2 - |F|$$
\end{theorem}
\begin{proof}
    When the coalition $F$ deviates from $\sigma$ to $\gamma$, the only nodes in the entire network $\mathcal{V}$ that experience a change in their individual payoff are the members of $F$ and their immediate neighbors in $D$. Nodes outside $F \cup D$ are unaffected. Therefore, the global payoff variation of the network can be partitioned as:$$\Delta\mu(\mathcal{V}) = \sum_{v \in F} \Delta\mu_v + \sum_{v \in D} \Delta\mu_v$$From the Global Variation Identity (Lemma 3.2), we also know that the global variation is exactly determined by the internal gains and the change in the internal cut:$$\Delta\mu(\mathcal{V}) = 2\left(\sum_{v \in F} \Delta\mu_v - \Delta P(F,F)\right)$$By equating the two expressions for $\Delta\mu(\mathcal{V})$, we can isolate the total payoff variation of the boundary $D$:$$\sum_{v \in F} \Delta\mu_v + \sum_{v \in D} \Delta\mu_v = 2\sum_{v \in F} \Delta\mu_v - 2\Delta P(F,F)$$$$\sum_{v \in D} \Delta\mu_v = \sum_{v \in F} \Delta\mu_v - 2\Delta P(F,F)$$To find the lower bound (the worst-case scenario) for the boundary $D$, we must minimize the positive term $\sum_{v \in F} \Delta\mu_v$ and maximize the subtractive term $2\Delta P(F,F)$.Because $F$ performs a strong deviation, every deviating node strictly improves its payoff by at least $1$. Thus, the aggregate minimum gain for the coalition is exactly its size:$$\sum_{v \in F} \Delta\mu_v \ge |F|$$Because the initial state $\sigma$ is a Nash Equilibrium, the Internal Cut Growth Constraint (Lemma 3.3) dictates that the increase in internal conflicts is strictly less than the number of deviating nodes ($\Delta P(F,F) < |F|$). Since we operate on a discrete unweighted graph, the maximum integer value for the internal cut growth is exactly $|F| - 1$:$$\Delta P(F,F) \le |F| - 1$$Substituting these discrete bounds into our equation for $D$ yields:$$\sum_{v \in D} \Delta\mu_v \ge |F| - 2(|F| - 1)$$$$\sum_{v \in D} \Delta\mu_v \ge |F| - 2|F| + 2$$$$\sum_{v \in D} \Delta\mu_v \ge 2 - |F|.$$
    This concludes the proof. 
\end{proof}

The above Theorem is relevant since it shows that the possible losses to the welfare produced by the coalition to the boundary set $D$ cannot exceed the number of elements of the coalition minus 2.

Moreover, we show that under the hypothesis of boundary-disjoint colors, introduced in Definition \ref{disjoint}, the total payoff induced to the boundary set $D$ by any strong coalition is non-decreasing.

\begin{theorem}[Boundary Disjoint Deviation]\label{bound_disj}
    Let $\sigma$ be a Nash Equilibrium. If a coalition $F$ performs a strong, boundary-disjoint deviation to $\gamma$, then no node in the boundary $D$ experiences a decrease in payoff. Furthermore, the total payoff of $D$ is strictly non-decreasing. Formally:
\begin{itemize}
    
\item[i.] $\forall v \in D, \Delta\mu_v(\gamma, \sigma) \ge 0$,
\item[ii.]$\sum_{v \in D} \Delta\mu_v(\gamma, \sigma) \ge 0$.
\end{itemize}
\end{theorem}
\begin{proof}
    The payoff of a node $v \in D$ strictly depends on the number of conflicts it has with its neighbors. Let $c_v \ge 0$ be the number of edges between $v$ and nodes in $F$ that share the same strategy (color) as $v$ under the initial configuration $\sigma$. When $F$ deviates to $\gamma$, by the definition of a boundary-disjoint deviation, no node in $F$ adopts the strategy of $v$. Therefore, the new number of conflicts between $v$ and $F$ is exactly $0$.The variation in payoff for the individual node $v$ is the difference between its old conflicts and its new conflicts:$$\Delta\mu_v(\gamma, \sigma) = c_v - 0 = c_v.$$Since the number of initial conflicts $c_v$ cannot be negative ($c_v \ge 0$), it follows trivially that $\Delta\mu_v(\gamma, \sigma) \ge 0$ for every individual node in the boundary. Consequently, the sum of their variations is also non-negative. 
\end{proof}

The previous findings on the effect of the strong coalitions on the social welfare suggest to extend to the general case, thus providing a theoretical framework to overcome the Nash individualistic traps. This result should allow a system to move from a static state to a richer metastable one, driving the whole system towards effective welfare  improvements. To demonstrate this result formally is not possible using the assumptions used along this paper, then we formulate a conjecture.

\begin{conjecture}[Monotone Social Welfare]
\label{thm:monotone}
Any strong deviation by a coalition $F$ from a sub-optimal (Nash) coloring strictly increases the  social welfare of the system ($\Delta \mu(\gamma, \sigma) > 0$).
\end{conjecture}

As said before, we are unable to formally prove that strong deviating coalitions induce a non-negative payoff variation on the frontier $D$ of the coalition itself ($\sum_{w \in D} \Delta\mu_w \ge 0.$), and, as a consequence a strict increase of the global welfare $\Delta \mu > 0$). 

Anyway, we were unable to find any counterexample by means of a computational extensive search, as it is reported in the Main Document.

\subsection{Metastability and Escape Thresholds} 

While Theorem \ref{3punto1} establishes that the global optimum is a SE, the system often resides in intermediate configurations stable against individual moves but vulnerable to collective ones. The persistence of these metastable Nash Equilibria is rooted in the structural inertia of the network.

According to Lemma \ref{variation_identity}, a transition toward higher welfare requires the aggregate individual gains to outweigh the internal reconfiguration cost:
\begin{equation}
\sum_{v \in F} \Delta \mu_v(\gamma, \sigma) > \Delta P(F, F).
\end{equation}

While Lemma \ref{lemma:conflicts_density} guarantees $\Delta P(F,F) < |F|$, showing escape is theoretically possible, the actual feasibility depends on the chromatic degrees of freedom within the subgraph $G[F]$, defined on the nodes of the subset $F$. If a coalition is too small relative to internal conflict density, no recoloring $\gamma$ satisfies $\Delta \mu_v \geq 1$ for all members simultaneously.

\begin{proposition}[Existence of Metastability Radius]\label{prop45}
Let $\sigma$ be a metastable coloring. Then there exists a well-defined minimum integer $m \geq 2$, the metastability radius of $\sigma$, such that $\sigma$ is $m$-rank resilient: no coalition $F$ with $|F| < m$ can perform a strong deviation from $\sigma$.
\end{proposition}

\begin{proof}
Since $\sigma$ is a NE, no coalition of size $|F| = 1$ can deviate by definition: $\sigma$ is at least 2-rank resilient. Since $\sigma$ is metastable (NE but not SE), there exists at least one coalition $F^* \subseteq V$ performing a strong deviation. Define:
\begin{equation}
M(\sigma) = \{ |F| : F \subseteq V, F \text{ performs a strong deviation from } \sigma \}
\end{equation}
$M(\sigma)$ is non-empty (by metastability) and contained in $\{2, \dots, N\}$ (by the NE property). Since it is a non-empty finite set of positive integers, it admits a well-defined minimum:
\begin{equation}
m = m(\sigma) := \min M(\sigma)
\end{equation}
By construction, no coalition of size strictly less than $m$ can perform a strong deviation, which is precisely the definition of $m$-rank resilience.
\end{proof}

Proposition \ref{prop45} establishes that evolution in complex interaction networks is not a continuous process but a discrete sequence of jumps, each requiring a coalition reaching or exceeding the critical mass $m$. Smaller coalitions are structurally incapable of breaking the equilibrium. This provides a formal foundation for the concept of critical mass in evolutionary biology \cite{Odum1969}: the minimum number of coordinated agents required to trigger a systemic transition.

Moreover, we notice that the metastability radius $m$ is not a fixed property but evolves dynamically as coalitions drive the network toward the global optimum. As the system approaches the climax, residual sub-optimality becomes increasingly localized, and smaller coalitions suffice. This predicts a monotone non-increasing behavior of $m$ along the trajectory.


\section{Biological Application and Pollination Network Modeling}

In this section, we describe the methodology used to translate a real ecosystem into a dynamic game-theoretic model. The objective is to empirically validate how the search for maximum global welfare, hereby referred to as \textit{climax}, is driven by symbiotic dynamics and the functional partitioning of ecological niches.


For our analysis, we utilized a global plant-pollinator interaction dataset provided in \cite{THEBAULT2020, ARROYO1985}, specifically focusing on the local network identified as Network ID 3 from experiments conducted in the high Andes of central Chile. This alpine environment is characterized by a contracted seasonality and limited energy resources, factors that exacerbate competitive dynamics among species. The system consists of 28 pollinator species acting as the agents, interacting with 41 plant species, representing the resources, for a total of 91 observed interactions. The taxonomical distribution of pollinators, reported in Table \ref{tab:pollinators}, reveals a predominance of \textit{Diptera} and \textit{Lepidoptera}.

\begin{table}[h]
\centering
\caption{Taxonomical distribution of pollinators (Network ID 3).}
\label{tab:pollinators}
\begin{tabular}{lc}
\hline
Order (Pollinator Order) & Number of Species \\ \hline
Diptera & 11 \\
Lepidoptera & 8 \\
Hymenoptera & 6 \\
Coleoptera & 1 \\
n.i. (Not identified) & 2 \\ \hline
\end{tabular}
\end{table}

A quantitative analysis reveals a state of high saturation, in fact nearly half of the plant species (48.8\%) are shared by multiple pollinators, indicating that system frustration is not merely a theoretical hypothesis but a structural condition arising from the contest for common resources.


We describe the ecosystem as an undirected and unweighted graph $G = (\V, \E)$, including the follows elements:
\begin{itemize}
    \item Nodes ($\V$): Represent the agents of the system, i.e., the species of pollinators (bees, flies, butterflies, beetles). Each species is treated as a rational agent aiming to maximize its own \textit{biological fitness}.
    \item Edges ($\E$): Represent trophic niche overlap. An edge $a_{v,w} = 1$ exists if and only if species $v$ and $w$ visit at least one common plant resource ($F_v \cap F_w \neq \emptyset$). Biologically, an edge indicates competitive pressure: if an edge exists, the two species are in conflict over the same food.
    \item Colors ($\sigma$): The concept of color is interpreted as the foraging strategy or functional niche. Assigning a color defines the ecological role of a species (e.g., preference for specific flowers, activity time, morphological traits).
\end{itemize}

In this biological mapping, we identify \textit{symbiotic guilds} as the physical manifestation of deviating coalitions $F$. While traditional competition models focus on individual displacement, our model shows that these guilds act as the primary drivers toward the climax state.

\begin{definition}[Biological Climax]
We define the \textit{biological climax} of an ecosystem $G$ as any strategy profile $\sigma^*$ that satisfies the global optimum condition:
\begin{equation}
\sigma_{Climax} \in \arg\max_{\sigma \in \mathcal{K}^N} \mu(\mathcal{V}, \sigma).
\end{equation}
In ecological terms, this represents the state of maximum niche differentiation. In this configuration, species have successfully partitioned available resources to minimize functional overlap, effectively mitigating the high costs of competition that arise when occupying identical niches.
\end{definition}

We establish that the biological fitness of each species $v$ is determined by its success in niche partitioning. Following the principle that competition between individuals in the same niche is significantly stronger than between individuals in different ones \cite{Adler2018}, our model rewards agents that differentiate their traits (colors). In our unweighted graph, the fitness $\mu_v(\sigma)$ is the number of competitors with whom the species has successfully differentiated its strategy:
\begin{equation}
\mu_v(\sigma) = \sum_{\substack{w \in \V \\ \sigma_w \neq \sigma_v}} a_{v,w}.
\end{equation}
By maximizing this value, species effectively reduce their niche overlap, transitioning from a state of taxonomic chaos to a stable, functionally integrated community.


To evaluate the optimality of the climax state, we conducted a Monte Carlo simulation with 100,000 random samples. We compared three states: the random distribution induced by stochastic noise, the taxonomical division for which ancestral state where species are grouped by phylogeny, and the climax, i.e. the social optimum reached via the max $k$-cut game.

\begin{figure}[ht!]
    \centering
    \includegraphics[width=0.8\textwidth]{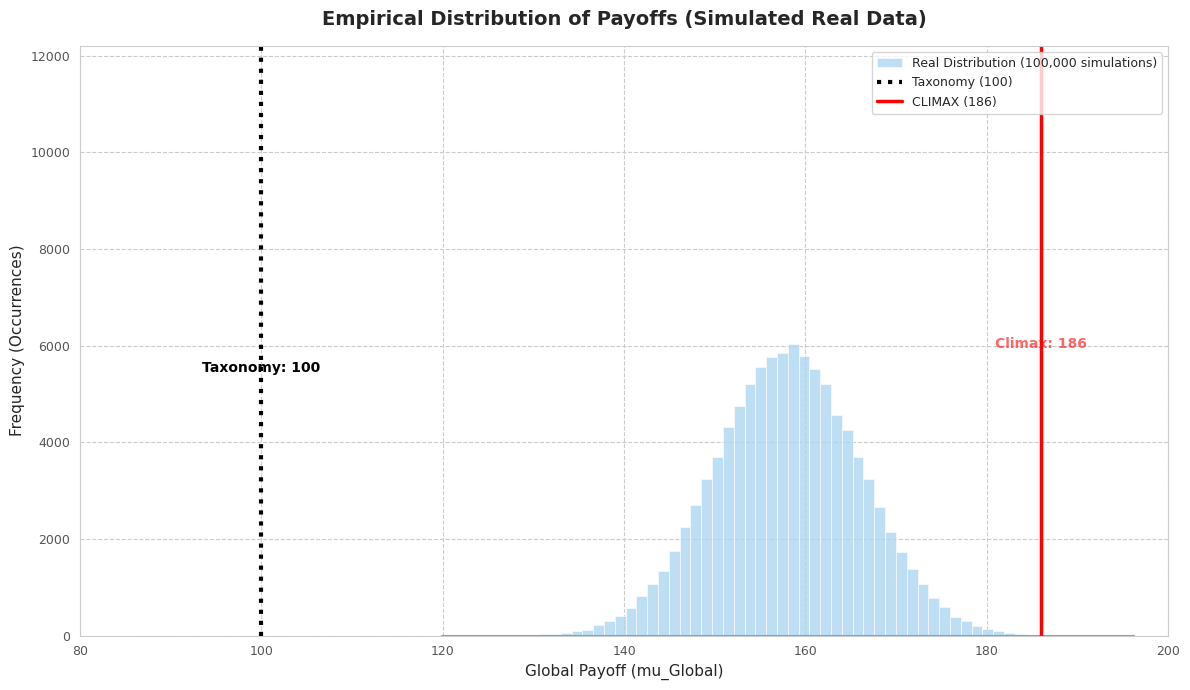}
    \caption{Statistical validation of the Biological Climax via Monte Carlo simulation. The blue histogram represents the empirical distribution of global payoffs obtained from 100,000 random niche assignments ($k=5$). The distribution approximates a Gaussian curve centered at a mean of $\approx 157$, which defines the background noise of a system driven purely by stochasticity. Two critical benchmarks are highlighted: (i) the taxonomic state of 100 (dashed black line) is located more than 7 standard deviations to the left of the mean and (ii) the biological climax  at 186 (solid red line) resides at the far right tail, exceeding the maximum value achieved in any of the 100,000 random trials, confirmed statistically by a p-value $< 10^{-5}$.}
\label{fig:statistical_validation}
\end{figure}

As shown in Figure \ref{fig:statistical_validation}, this analysis reveals the exceptionality of the optimized state, establishing a clear distinction between stochastic noise, ancestral inheritance, and functional optimization.

\begin{table}[h]
\centering
\caption{Final statistical validation evidence (Network ID 3, $k = 5$).}
\label{tab:stats}
\begin{tabular}{ll}
\hline
Metric & Value \\ \hline
Mean Payoff (Random) & 156.8269 \\
Standard Deviation ($\sigma$) & 7.9030 \\
Z-Score CLIMAX (186) & +3.6914 $\sigma$ \\
Z-Score TAXONOMY (100) & --7.1906 $\sigma$ \\
P-Value CLIMAX & $1.12 \times 10^{-4}$ \\
Percentile CLIMAX & 100.0000\% \\ \hline
\end{tabular}
\end{table}

The interpretation of the results shown in Figure \ref{fig:statistical_validation} and Table \ref{tab:stats} data is profound giving rise to the following main findings: $(i)$ Inefficiency of phylogeny: the taxonomical state of 100 sits more than 7 standard deviations below the mean, proving that niche conservatism represents a state of extreme biological suffering and inefficiency. $(ii)$ Excellence of climax: the climax state of 186 achieved a payoff higher than any of the 100,000 random simulations. The probability of this state occurring by chance is approximately 0.01\%. $(iii)$ Validation of the strong equilibrium: the rarity of the 186 value proves that the strong equilibrium is not a random aggregate but the result of a directional evolutionary drive toward maximum resilience and minimum frustration, making it resilient to stochastic fluctuations and coordinated opportunistic shifts.

The transition from taxonomical chaos to functional climax was modeled through a series of symbiotic guild formations. Each step follows the monotone optimization principle, expressed by Conjecture \ref{thm:monotone}, for which coalitional deviations increase both individual and global fitness.

\begin{table}[h]
\centering
\caption{Chronology of the evolutionary climb via symbiotic guilds ($r > 1$).}
\label{tab:chronology}
\begin{tabular}{clcc}
\hline
Step & Deviating Coalition ($F$) & Size ($r$) & Payoff ($\mu_{Global}$) \\ \hline
0 & Ancestral State (Taxonomy) & - & 100 \\
1 & \textit{Phulia nymphula, Sarcophagidae n.i., etc.} & 4 & 130 \\
2 & \textit{Anthidium sp., Hypsochila wagenknecti, etc.} & 4 & 146 \\
3 & \textit{Cynthia terpsichore, Diptera n.i., etc.} & 4 & 166 \\
4 & \textit{Yramea modesta, Megachile semirufa, etc.} & 4 & 180 \\
5 & \textit{Bombus dahlbomii, Stenanthidium espinosai, etc.} & 3 & \textbf{186} \\ \hline
\end{tabular}
\end{table}

At the climax the system reaches a highly efficient state, as shown by the 93 cuts edges out of 98, but it includes 5 residual conflicts remain. These welding points are topologically insoluble without causing negative externalities to the rest of the community, confirming that real biodiversity is a compromise between theoretical perfection and network constraints.

\begin{figure}[ht!]
    \centering
    \includegraphics[width=0.85\textwidth]{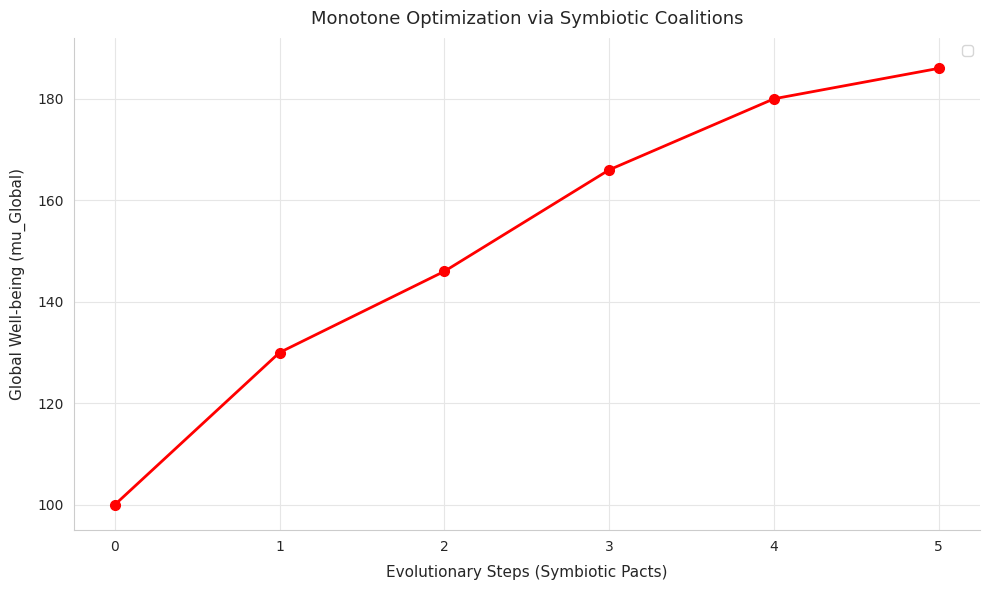}
    \caption{Monotone optimization of social welfare via symbiotic pacts. The red curve illustrates the global payoff $\mu_{Global}$ as a function of discrete evolutionary steps. Starting from the suboptimal taxonomic state (Step 0, Payoff = 100), the system undergoes five coordinated shifts induced by symbiotic pacts, thus reaching the biological climax (Payoff = 186).}
    \label{fig:monotone_optimization}
\end{figure}

The strictly ascending trajectory of the system's fitness is captured in Figure \ref{fig:monotone_optimization}, which records the global social welfare across successive evolutionary stages. This monotonic progression empirically validates our theoretical result that coordinated symbiotic deviations act as a one-way ratchet toward higher systemic efficiency. Each step corresponds to a strong deviation by a functional guild that strictly improves the utility of its members while simultaneously increasing the total network cut. The absence of any downward fluctuations provides empirical proof of the exact potential game nature of the max $k$-cut game, where local coordination inherently aligns with global optimization. The curve asymptotically approaches the state where the system achieves maximum functional integration and transitions into a resilient strong equilibrium. From a dynamical systems perspective, any intermediate steps towards the climax can be seen as metastable states, where their suboptimality is transitory and is step-by-step overwhelmed by the emergence of successive coalitions.

The evolutionary trajectory detailed in Table \ref{fig:monotone_optimization} provides a direct empirical validation of the Metastability Radius established in Proposition \ref{prop45}. The transition from the sub-optimal Ancestral State (Step 0) to the first improved configuration (Step 1) was triggered by a coalition of exactly $r=4$ species. This indicates that the taxonomic state functioned as a 4-rank resilient trap: any individual move or smaller coalition would have been unable to overcome the structural inertia of the network. Each successive step in the climb, characterized by guild sizes of $r=3$ or $r=4$, represents the minimum critical mass predicted by our model. These guilds are the physical manifestation of the escape from a metastable state, where the collective synergy of the species was sufficient to outweigh the internal reconfiguration costs $\Delta P(F, F)$ that had previously locked the system in a state of high competition and low efficiency.

\section{Conclusion}
In this Supplementary Material document, we have analytically demonstrated Lemmas, Propositions and Theorems leading to demonstrate that in the context of anti-coordination games on unweighted networks, global efficiency and coalitional stability are perfectly aligned. Our main result (Theorem \ref{3punto1}) proves that no group of players, regardless of their size or coordination capabilities, can improve their collective standing once the system reaches a social optimum.

Moreover, we report the details of the application to the high-Andes pollination network, thus confirming that evolution acts as a one-way ratchet: symbiotic coalitions drive the system away from the inefficiency of taxonomic state toward a climax. 

This mathematical framework explains why mature ecosystems exhibit such high resilience against internal opportunistic shifts, identifying symbiosis not just as a contingency but as a systemic catalyst for global optimization.





%